\documentclass[12pt,preprint]{aastex}

\usepackage{graphicx}

\slugcomment{The Astronomical Journal, in press}

\shorttitle{Elst Pizarro}
\shortauthors{Jewitt et al.}

\begin{document}

%\title{Hubble Space Telescope Observations of Main-Belt Comet 133P/Elst-Pizarro
%}
\title{HUBBLE SPACE TELESCOPE INVESTIGATION OF MAIN-BELT COMET 133P/ELST-PIZARRO
}
\author{David Jewitt$^{1,2}$, Masateru Ishiguro$^{3,1}$, Harold Weaver$^4$, Jessica Agarwal$^5$, Max Mutchler$^6$ and Steven Larson$^7$
}
\affil{$^1$Dept.~of Earth, Planetary and Space Sciences,
University of California at Los Angeles, \\
595 Charles Young Drive East, 
Los Angeles, CA 90095-1567\\
$^2$Dept.~of Physics and Astronomy,
University of California at Los Angeles, \\
430 Portola Plaza, Box 951547,
Los Angeles, CA 90095-1547\\
$^3$ Dept.~Physics and Astronomy, Seoul National University, Gwanak, Seoul 151-742, Republic of Korea \\
$^4$ The Johns Hopkins University Applied Physics Laboratory, 11100 Johns Hopkins Road, Laurel, Maryland 20723  \\
$^5$ Max Planck Institute for Solar System Research, Max-Planck-Str. 2, 37191 Katlenburg-Lindau, Germany
$^6$ Space Telescope Science Institute, 3700 San Martin Drive, Baltimore, MD 21218 \\
$^7$ Lunar and Planetary Laboratory, University of Arizona, 1629 E. University Blvd.
Tucson AZ 85721-0092 \\
}

\email{jewitt@ucla.edu}

\begin{abstract}
We report new observations of the prototype main-belt comet (active asteroid) 133P/Elst-Pizarro taken at high angular resolution using the Hubble Space Telescope.  The object has three main components; a) a point-like nucleus, b) a long, narrow antisolar dust tail and c) a short, sunward anti-tail.   There is no resolved coma. The nucleus has a mean absolute magnitude $H_V$ = 15.70$\pm$0.10 and a lightcurve range $\Delta V $ = 0.42 mag., the latter corresponding to projected dimensions 3.6$\times$5.4 km (axis ratio 1.5:1), at the previously measured geometric albedo of 0.05$\pm$0.02.  We explored a range of continuous and impulsive emission models to simultaneously fit the measured surface brightness profile,  width and position angle of the antisolar tail.  Preferred fits invoke protracted emission, over a period of 150 days or less, of dust grains following a differential power-law size distribution with index 3.25 $\le q \le$ 3.5 and having a wide range of sizes.   Ultra-low surface brightness dust projected in the sunward direction is a remnant from emission activity occurring in previous orbits, and consists of the largest ($\ge$cm-sized) particles.   Ejection velocities of one micron-sized particles are comparable to the $\sim$ 1.8 m s$^{-1}$ gravitational escape speed of the nucleus, while larger particles are released at speeds less than the gravitational escape velocity.  The observations are consistent with, but do not prove, a hybrid hypothesis in which  mass loss is driven by gas drag from the sublimation of near-surface water ice, but escape is aided by centripetal acceleration from the rotation of the elongated nucleus.  No plausible alternative hypothesis has been identified. 
%$\theta_{\odot}$

\end{abstract}


\section{INTRODUCTION}

In the traditional view, asteroids are rocky bodies formed inside the snow line of the Sun's proto-planetary disk while comets are ice-containing objects formed outside it. Recently, several hybrid objects having orbits interior to Jupiter's and the dynamical properties of asteroids, have been found to exhibit comet-like mass-loss (Hsieh and Jewitt 2006). Known as main belt comets (MBC) or, more generally, active asteroids (Jewitt 2012), these objects are candidate ice-bearing asteroids. Asteroid ice is of potentially far-reaching scientific interest.  Most importantly, the outer regions of the asteroid belt may have supplied some Terrestrial water and other volatiles, including the biogenic precursor molecules up to and including amino acids (Mottl et al.~2007). In the modern solar system, there is no known dynamical path linking stable orbits in the asteroid belt to the classical comet reservoirs in the Kuiper belt or Oort cloud. Instead, if ice exists in the asteroid main belt, it is likely to be primordial, trapped either at the formation epoch, or in some early, chaotic phase of solar system history (Levison et al.~2009). 

About a dozen examples of active asteroids are currently known, but they do not all contain ice.  Data in-hand already clearly show the action of several different processes in causing mass-loss, ranging from asteroid-asteroid impact, to rotational break-up, to thermal fracture and more (Jewitt 2012).  Only for two objects, 133P/Elst-Pizarro (Hsieh et al.~2004, Hsieh and Jewitt 2006, Hsieh et al.~2010) and 238P/Read (Hsieh and Jewitt 2006, Hsieh et al.~2011), is the evidence for water sublimation reasonably strong.  The most compelling evidence is the repeated activity exhibited by both bodies. This is naturally explained by the seasonal sublimation of near-surface ice (including effects due to the orbital eccentricity as well as the obliquity of the nucleus, and possible self-shadowing by surface topography) but  is difficult or impossible to reconcile with the other mass-loss mechanisms so far envisioned.  No gas has yet been spectroscopically detected in any of the active asteroids, but the available limits to gas production are consistent with the very low ($<$1 kg s$^{-1}$) mass-loss rates inferred from dust (Jewitt 2012).  The evidence for ice sublimation thus remains indirect but no other plausible explanation for the observations, in particular for the seasonal recurrence of the activity, is known.

In this paper we describe the first high angular resolution, time-resolved images of 133P obtained using the Hubble Space Telescope. Our observations were triggered by the report of new activity on UT 2013 June 4, the fourth such episode of activity in 133P observed since 1996 (Hsieh et al.~2013).  We use the images to examine in detail the spatial and temporal properties of the mass-loss in the near-nucleus environment and to test the viability of the sublimating ice hypothesis.

\section{OBSERVATIONS} 
We used two consecutive orbits of Target-of-Opportunity time (General Observer program number 13005) to observe 133P on UT 2013 July 10.  Consecutive orbits were secured in a deliberate attempt to observe time-dependent changes in the inner coma, as might be induced by the rotation of the underlying nucleus at period 3.471$\pm$0.001 hr (Hsieh et al.~2004).  We obtained a total of 12 images with the WFC3 camera (Dressel 2010). The 0.04\arcsec~pixels of WFC3 each correspond to about 60.2 km at the distance of 133P, giving a Nyquist-sampled spatial resolution of about 120 km.  All observations were taken using the F350LP filter. This very broad filter (full-width-at-half-maximum is 4758\AA, while the effective wavelength for a solar-type (G2V) source is 6230\AA) provides maximum sensitivity to faint sources at the expense of introducing some uncertainty in the transformation to standard astronomical filter sets.  We used the HST exposure time calculator to convert the measured count rate into an effective V magnitude, finding that a V = 0 G2V source gives a count rate of 4.72 $\times$10$^{10}$ s$^{-1}$ within a 0.2\arcsec~radius photometry aperture.

The observational geometry of the HST observations is summarized in Table (\ref{geometry}).

\section{DERIVED PROPERTIES OF 133P}
\begin{figure}
\epsscale{0.9}
\begin{center}
%\plotone{Figure_1.ps}
%\includegraphics[width=0.75\textwidth, angle =270 ]{Figure_1.pdf}
\includegraphics[width=0.75\textwidth]{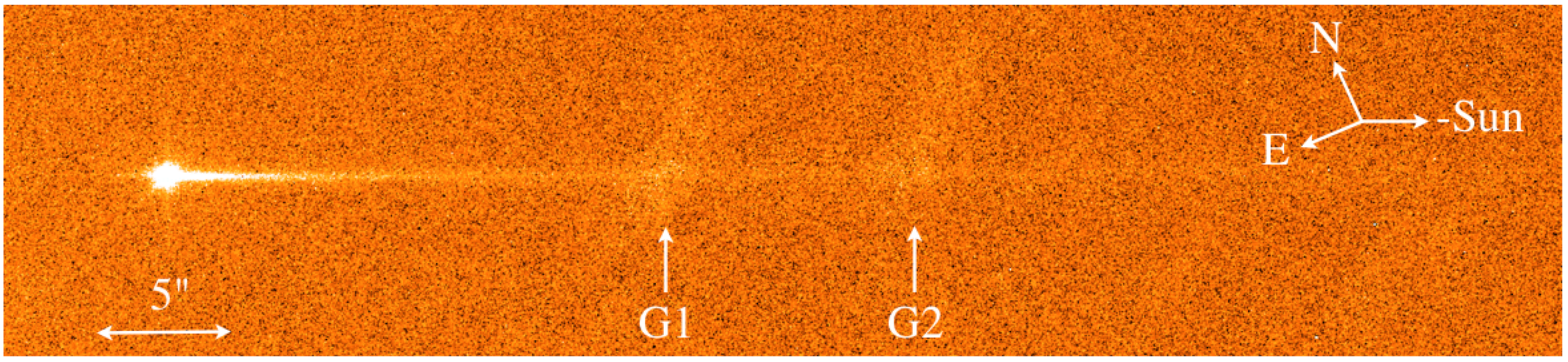}
\caption{Composite F350LP image of 133P having total integration time 3984 seconds.   A long, thin dust tail extends from the nucleus, visible at left. G1 and G2 mark field galaxies imperfectly removed from the data.  The antisolar direction is marked.  This is indistinguishable from the direction of the projected negative velocity vector at the resolution of the Figure (see Table \ref{geometry}). \label{image}
} 
\end{center} 
\end{figure}

Figure (\ref{image}) shows the drizzle-combination of all 12 exposures.  Cosmic rays and most background objects have been successfully removed from Figure (\ref{image}) except for residual signals from spatially extended objects (galaxies), some of which cross the tail obliquely ($G1$ and $G2$ in the Figure).  133P shows its characteristic point-like nucleus, while a thin, straight dust tail extends $>$60\arcsec~(90,000 km) to the edge of the field of view along position angle 247.1\degr$\pm$0.4\degr.  The position angles of the projected antisolar direction and the negative velocity vector were 246.8\degr~and 248.7\degr, respectively.  These directions are so close as to be indistinguishable in Figure \ref{image}.

\subsection{Nucleus}
\label{nucleus}
\begin{figure}
\epsscale{1.0}
\begin{center}
\plotone{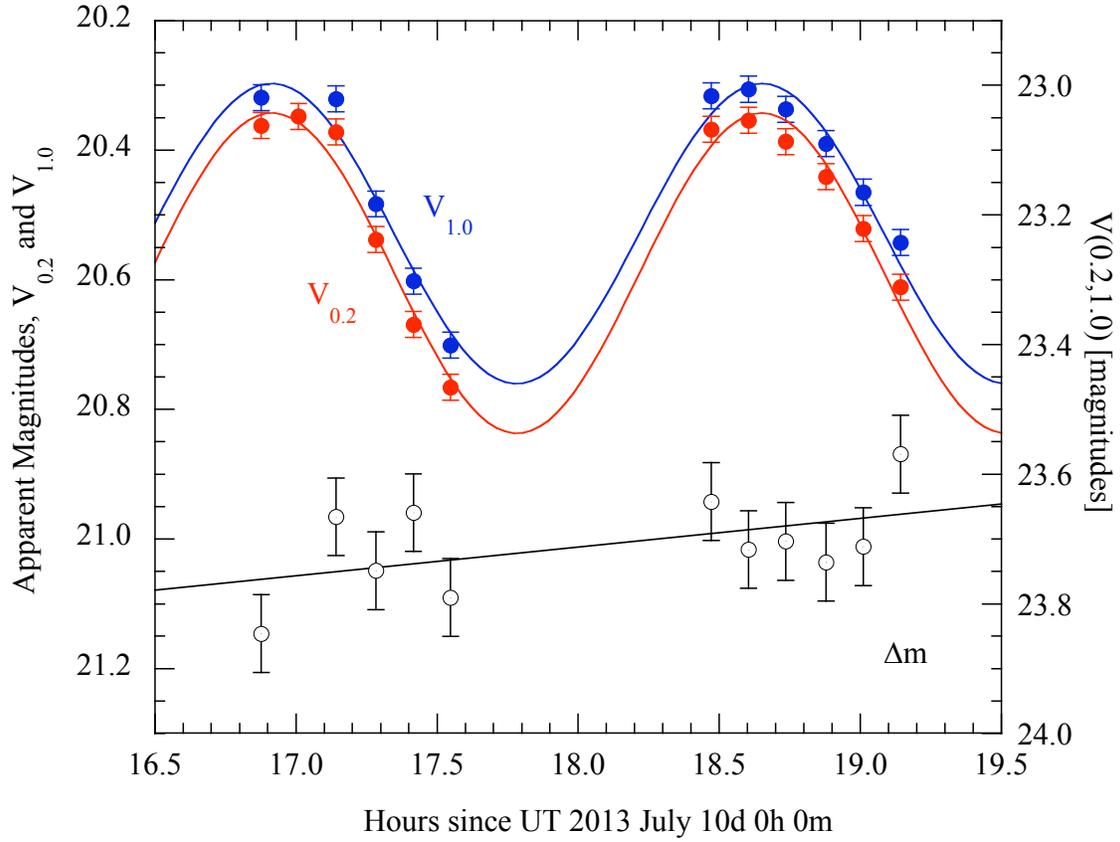}
\caption{Rotational lightcurve of 133P showing V$_{0.2}$ as red circles and V$_{1.0}$ as blue circles.  Red and blue lines mark the best-fit sinusoids and are added to guide the eye.  The magnitude of the material between the two apertures is also shown (black empty circles and right-hand vertical axis).  The black straight line is a least-squares fit to the data.  \label{lightcurve}
} 
\end{center} 
\end{figure}

The light from the central region is dominated by the bright and point-like nucleus.  Figure  (\ref{lightcurve}) shows aperture photometry measured from the individual images, after cleaning by hand for the removal of cosmic rays and artifacts.  To perform the latter, we first subtracted the median image from each individual image, in order to make visible cosmic rays and artifacts otherwise hidden in the bright near-nucleus region.  Then we removed cosmic rays one by one, using digital interpolation to replace affected pixels with the means of their surroundings.  Lastly, we added back the median image in order to recover the total signal.  For photometry, we used apertures 5 pixels (0.2\arcsec) and 25 pixels (1.0\arcsec) in radius, with sky subtraction determined from the median signal computed within a concentric annulus having inner and outer radii 25 and 50 pixels (2.0\arcsec), respectively.  We refer to magnitudes from these apertures as $V_{0.2}$ and $V_{1.0}$, respectively.  The use of such small apertures is enabled by the extraordinary image quality and pointing stability of the Hubble Space Telescope.  Photometry with the 0.2\arcsec~aperture samples primarily the nucleus, with only a small contribution from the surrounding dust.    
Photometry with the 1.0\arcsec~aperture includes both nucleus and dust, and is useful to compare with ground-based data in which atmospheric seeing precludes the use of sub-arcsecond apertures.    Figure  (\ref{lightcurve})  shows a clear temporal variation in both $V_{0.2}$ and $V_{1.0}$.    

The photometry and the physical properties of the comet are connected by the inverse square law of brightness, written

\begin{equation}
p_V \Phi(\alpha) C_n = 2.25\times10^{22} \pi R^2 \Delta^2 10^{-0.4(m - m_{\odot})}
\label{inversesq}
\end{equation} 

\noindent where R, $\Delta$ and $\alpha$ are from Table (\ref{geometry}), $\Phi(\alpha)$ is the phase function correction and $m_{\odot}$ = -26.75 is the apparent V magnitude of the Sun (Drilling and Landolt 2000).  The physical properties are the V-band geometric albedo, $p_V$, and the effective cross-section of the nucleus, $C_n$.   The albedo of 133P is $p_V$ = 0.05$\pm$0.02 (Hsieh et al.~2009).  For the correction to 0\degr~phase angle we use the measured (H,G) phase function from Figure 3 of Hsieh et al.~(2010), which gives 1.1$\pm$0.1 mag., corresponding to $\Phi(18.7\degr)$ = 0.36$\pm$0.03.  

The 0.2\arcsec~aperture gives our best estimate of the nucleus brightness, with only minimal contamination from the near-nucleus dust.  We estimate that the mean apparent magnitude is V = 20.55$\pm$0.05 (Table \ref{photometry}).  The resulting absolute magnitude (at R = $\Delta$ = 1 AU and $\alpha$ = 0\degr) is $H_V$ = 15.70$\pm$0.10. Substituting into Equation (\ref{inversesq}) we obtain the nucleus cross-section $C_n$ = 15$\pm$5 km$^2$ and the equivalent circular radius  $r_n$ = ($C_n/\pi)^{1/2}$ = 2.2$\pm$0.5 km.  The uncertainties on both numbers are dominated by uncertainty in $p_V$, itself a product of phase-function uncertainty. The full range of the measured lightcurve in Figure (\ref{lightcurve}) is $\Delta V$ = 0.42 mag.     If attributed to the rotational variation of the cross-section of an $a \times b$ ellipsoid, then $a/b$ = 10$^{0.4 \Delta V}$ = 1.5, with $a \times b$ = 2.7 $\times$ 1.8 km.  Formally, the axis ratio derived from the lightcurve is a lower limit to the true value because the HST sampling missed minimum light and because we observe only the projection of the true nucleus axis ratio into the plane of the sky.  In practice, no larger $\Delta V$ has been reported for 133P (Hsieh et al.~2004, 2009, 2010, Bagnulo et al.~2010); we suspect that 133P is viewed from a near-equatorial perspective and that $a/b$ = 1.5 is close to the true nucleus axis ratio.  

The minimum density needed to ensure that the material at the tips of a prolate body in rotation about its minor axis is gravitationally bound is approximately 

\begin{equation}
\rho_{n} \sim 1000 \left(\frac{3.3}{P}\right)^2 \left[\frac{a}{b}\right]
\label{density}
\end{equation}

\noindent where $P$ is the rotation period in hours (Harris 1996).  Substituting $P$ = 3.5, ($a/b$) = 1.5, we find $\rho_n \sim$ 1300 kg m$^{-3}$, identical to a value determined from ground-based data (Hsieh et al.~2004).  The relation between the derived density and the true bulk density of 133P is uncertain, pending determination of the strength properties of the nucleus.

\subsection{Dust Tail and Near-Nucleus Environment}
We also plot in Figure (\ref{lightcurve}) the effective magnitude of the material projected between the 0.2\arcsec~and 1.0\arcsec~annuli, computed from

\begin{equation}
V(0.2,1.0) = -2.5\times \log_{10}\left[10^{-0.4 V_{0.2}} -  10^{-0.4 V_{1.0}} \right].
\label{mcoma}
\end{equation}

\noindent Evidently, the coma in the 0.2\arcsec~to 1.0\arcsec~annulus is steady or  slightly increasing in brightness with time and  does not share the temporal variability exhibited by the nucleus itself.  This is most likely because spatial averaging inhibits our ability to detect rotational modulation in the dust ejection rate.  To see this,  assume that the dust ejection speeds are comparable to the nucleus gravitational escape speed, $V_e \sim$ 1 m s$^{-1}$.  The time taken for dust to cross the photometry annulus is $\tau_c \sim \delta r/V_e$, where $\delta r \sim 10^3$ km is the linear distance between the 0.2\arcsec~and 1.0\arcsec~radius apertures.  We find $\tau_c \sim 10^6$ s, far longer than the $P \sim 10^4$ s nucleus rotation period.  Therefore, the photometry annulus should contain dust produced over  $\tau_c/P \sim$ 10$^2$ nucleus rotations, effectively smoothing out any rotation-dependent modulation of the signal.

\begin{figure}
\epsscale{0.9}
\begin{center}
\plotone{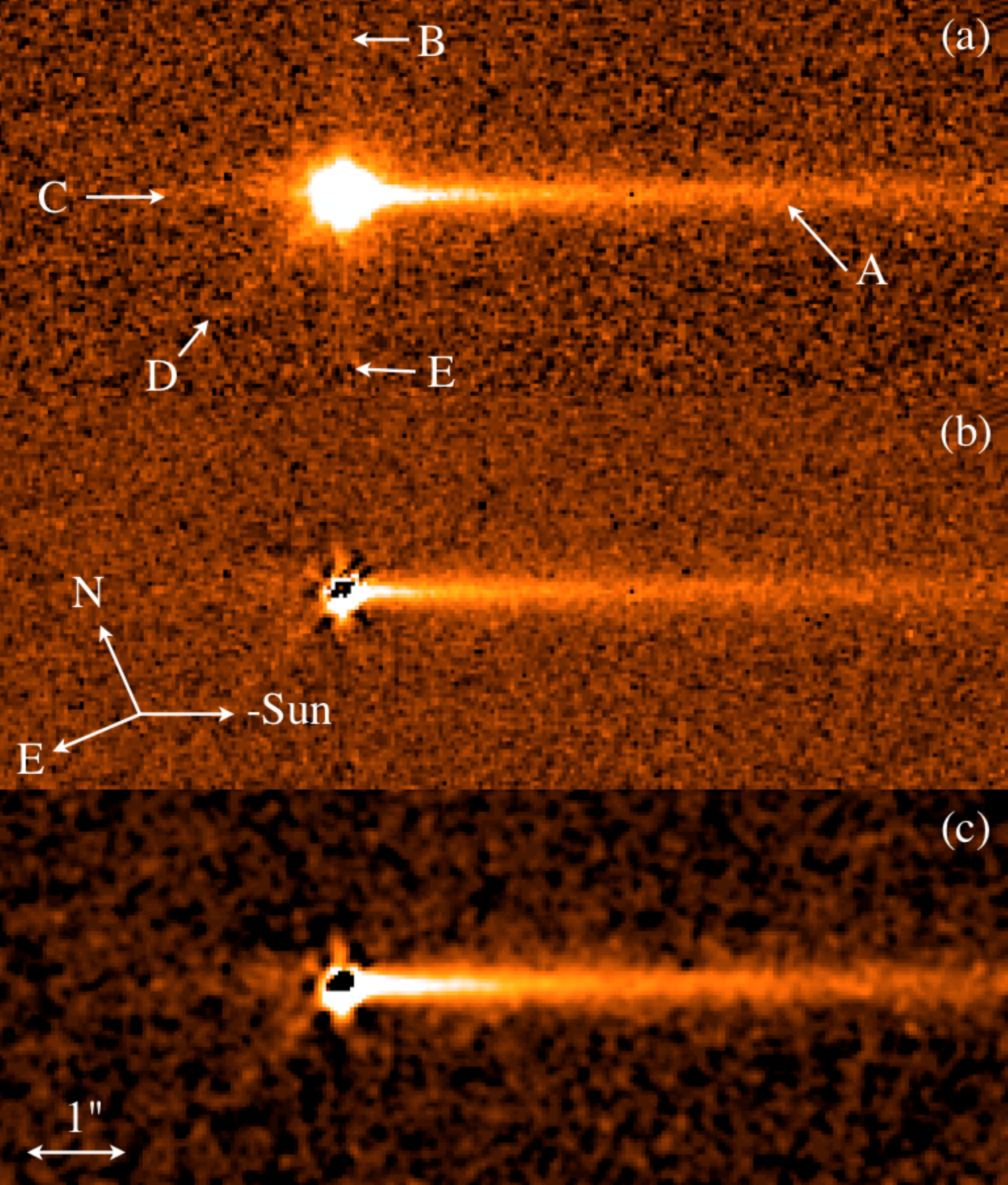}
\caption{(a) Near-nucleus region of the composite F350LP image from Figure \ref{image}.   Of the features labeled A-E, B and E are diffraction spikes, D is a charge-transfer efficiency artifact in the CCD and only A and C are true dust features associated with 133P. (b) Same image but with a centered, scaled model of the Hubble PSF subtracted (c) Gaussian-convolved and heavily-stretched version of (b) to show residual emission at locations of C and D.  A 1\arcsec~scale bar is shown.  \label{difference}
} 
\end{center} 
\end{figure}

Figure (\ref{image}) hints at the presence of several near-nucleus dust structures.  To examine these, we need to consider diffraction in the telescope optics and other artifacts.  For this purpose, we computed point-spread function (PSF) models using the ``TinyTim'' software package (Krist et al.~2011).   We assumed a G2V stellar spectrum and centered and scaled the model PSF to the data using photometry extracted from a circular aperture of projected radius 2 pixels.  The optimum scaling and centering were rendered slightly uncertain by the near-nucleus morphology of 133P.  We experimented with different scale factors and centers, finding that the results presented here are stable with respect to these factors.  The resulting difference image (Figure \ref{difference}b) successfully removes the diffraction spikes B and E. Feature D is an artifact caused by imperfect charge-transfer efficiency in the WFC3 detector and not modeled in the TinyTim software (c.f. Weaver et al.~2010).    Another diffraction spike underlies feature C, but residual emission survives in the difference image.   A lightly-smoothed version of Figure \ref{difference}b is shown in Figure \ref{difference}c, to emphasize the faint, surviving feature C.  We conclude that the major, measurable components of the image of 133P are 1) the nucleus, 2) the main tail, A, to the west of the nucleus and 3) a stubby, sunward tail, C, to the east.  The nucleus and the main tail have been defining signatures of 133P in all previous observations of 133P when in its active state (Hsieh et al.~2004).  The sunward tail has not been previously reported.

No coma is visually apparent in the HST data.  The 1.6 pixel (0.064\arcsec) full-width-at-half maximum of the TinyTim model PSF is very close to the 2.2 pixels (0.086\arcsec) FWHM of the combined 133P image, measured in the same way.  The absence of a coma is a clear indicator of the extremely low velocities with which dust particles are ejected from 133P.  A crude estimate of these velocities is obtained by considering the turn-around distance of a particle ejected towards the Sun at speed $v$, which is given by $X_R = v^2/(2\beta g_{\odot})$ (Jewitt and Meech 1987).  Dimensionless factor $\beta$ is a function of particle size, approximately given by $\beta \sim 1/a_{\mu m}$, where $a_{\mu m}$ is the grain radius expressed in microns.  Substitution gives

\begin{equation}
v^2 a_{\mu m} = 2 g_{\odot} X_R
\end{equation}

\noindent as a constraint on the particles that can be ejected from the nucleus of 133P.  Substituting $X_R <$ 150 km (0.1\arcsec) and $g_{\odot}$ = 8$\times$10$^{-4}$ m s$^{-2}$ for $R$ = 2.7 AU, we obtain $v^2 a_{\mu m} <$ 240 m$^2$ s$^{-2}$.   For instance, with $a_{\mu m}$ = 1, we find $v <$ 15 m s$^{-1}$ while millimeter-sized particles, $a_{\mu m}$ = 1000, have $v <$ 0.5 m s$^{-1}$.   The absence of a resolved coma therefore requires that particle ejection speeds be far smaller than the expected thermal speed of gas molecules ($V_g \sim$ 500 m s$^{-1}$) at this heliocentric distance.  We will reach a similar conclusion, later, in consideration of the narrow width of the tail of 133P. For comparison, a short-period comet at similar distance might have $X_R \sim$3$\times$10$^7$ m (Jewitt and Meech 1987), giving $v^2 a_{\mu m} \sim $ 5$\times$10$^4$ m$^2$ s$^{-2}$.  For a given particle size, $a_{\mu m}$, the dust from 133P is ejected sunward about (5$\times$10$^4$/240)$^{1/2} \sim$ 15 times or more slowly than from the short period comet.  Non-detection of a coma is our first indication of the very low dust speeds in 133P.

We further examined the near-nucleus region in search of dust structures that might be carried around by the rotation of the nucleus.  For this purpose, we subtracted the median image from the individual images  in order to enhance temporal and spatial changes. We found by experiment that centering uncertainties as small as $\pm$0.1 pixel (0.004\arcsec) had a large influence in the resulting difference images, to the extent that we could not reliably identify any evidence for a rotating pattern of emission.

\section{DUST MODELS}

\subsection{Order of Magnitude Considerations\label{OOM}}
The observed length of the main tail is $\ell_T \gtrsim$ 90,000 km.  This is a lower limit to the true length both because the tail extends beyond the field of view of HST and because we observe only the projection of the tail in the plane of the sky.  The time taken for radiation pressure to accelerate a grain over distance $\ell_T$ is $\tau_{rp} \sim (2 \ell_T/\alpha_g)^{1/2}$, where $\alpha_g$ is the grain acceleration.  We write $\alpha_g = \beta g_{\odot}$, where $g_{\odot}$ = 8 $\times$ 10$^{-4}$ m s$^{-2}$ is the gravitational acceleration to the Sun at the heliocentric distance of 133P.  Substituting $\beta \sim 1/a_{\mu m}$, we find that the radiation pressure timescale for the main tail is $\tau_{rp} \sim 5 \times 10^5 a_{\mu m}^{1/2}$ s.  A 1 $\mu m$ grain would be swept from the tail in 5$\times$10$^5$ s ($\sim$6 days) while a 100 $\mu m$ grain would take $\sim$2 months to be removed.  Equivalently, if we suppose that the main tail consists of particles ejected no earlier than the date on which the present active phase of 133P was discovered, UT 2013 June 4 (Hsieh et al. 2013), then the particle size implied by the length of the tail is $a_{\mu m} \sim$ 50 $\mu m$.   This is a lower limit to the particle size because larger particles (smaller $\beta$) would not have reached the end of the tail, and because the tail is certainly older than the June 4 date of discovery.  While clearly very crude, this estimate shows that the particles in the tail are larger than those typically observed at optical wavelengths and provides a basis for comparison with more sophisticated calculations, described in Section \ref{details}.

The width of the antisolar tail was measured from a series of profiles extracted perpendicular to the tail.  We binned the data parallel to the tail in order to increase the signal-to-noise ratio of the measurement, with larger bin sizes at the distant, fainter end of the tail than near the nucleus.  The width was estimated from the FWHM of the profiles and is shown in Figure (\ref{width}).  The Figure shows estimated statistical uncertainties on the widths, but these underestimate the total uncertainties, which include significant systematic effects due to background sources.  The trail width increases from $\theta_w$ = 0.2\arcsec~(width $w_T \sim$ 300 km) near the nucleus to $\theta_w$ = 0.5\arcsec~($w_T \sim$ 800 km) 13\arcsec~west along the tail.  The very faint sunward tail was also measured, albeit with great difficulty (Figure \ref{width}).

%%%
At the time of observation, Earth was only 0.54\degr~below the orbital plane of 133P (Table \ref{geometry}), so that $w_T$ largely reflects the true width of the tail perpendicular to the orbit.  The width of the dust tail is related to $V$, the component of the dust ejection velocity measured perpendicular to the orbital plane, by $w_T$ = 2 $V \delta t$, where $\delta t$ is the time elapsed since release from the nucleus, provided $\delta t \ll$ the orbital period of 133P.  The factor of 2 arises because particles are ejected both above and below the orbital plane. We write $V = V_1/a_{\mu m}^{1/2}$, where $V_1$ is the perpendicular ejection velocity of an $a_{\mu m}$ = 1 particle.  Assuming that the dust motion parallel to the orbit plane is determined by radiation pressure acceleration, we may write the distance of travel from the nucleus as $\ell_T = (1/2) \beta g_{\odot} \delta t^2$.  Again, setting $\beta \sim$ 1/$a_{\mu m}$, we can eliminate $\delta t$ between these equations to find

\begin{equation}
V_{1} = \left(\frac{g_{\odot} w_T^2}{8 \ell_T}\right)^{1/2}
\label{vperp1}
\end{equation}

\noindent  Equation (\ref{vperp1}) shows that $w_T \propto \ell_T^{1/2}$, broadly consistent with the trend of the measurements in Figure (\ref{width}).  We fitted Equation (\ref{vperp1}) to the width vs.~length data plotted  in Figure (\ref{width}), finding $V_{1}$ = 1.8 m s$^{-1}$ and hence 

\begin{equation}
V(a) = \frac{1.8}{a_{\mu m}^{1/2}}~[m~s^{-1}]
\label{vperp}
\end{equation}

\noindent This estimate is made assuming that the tail lies in the plane of the sky, and hence that the measured length is not foreshortened.  Equation (\ref{vperp}) gives extraordinarily small velocities compared to the sound speed in gas ($V_g \sim$ 500 m s$^{-1}$), and implies $V(1~mm) \sim$ 6 cm s$^{-1}$ for 1 mm particles and $V(1~cm) \sim$ 1.8 cm s$^{-1}$ for centimeter-sized grains.  We will return to this point in Section \ref{discussion}.

%%%%

\begin{figure}
\epsscale{0.9}
\begin{center}
\plotone{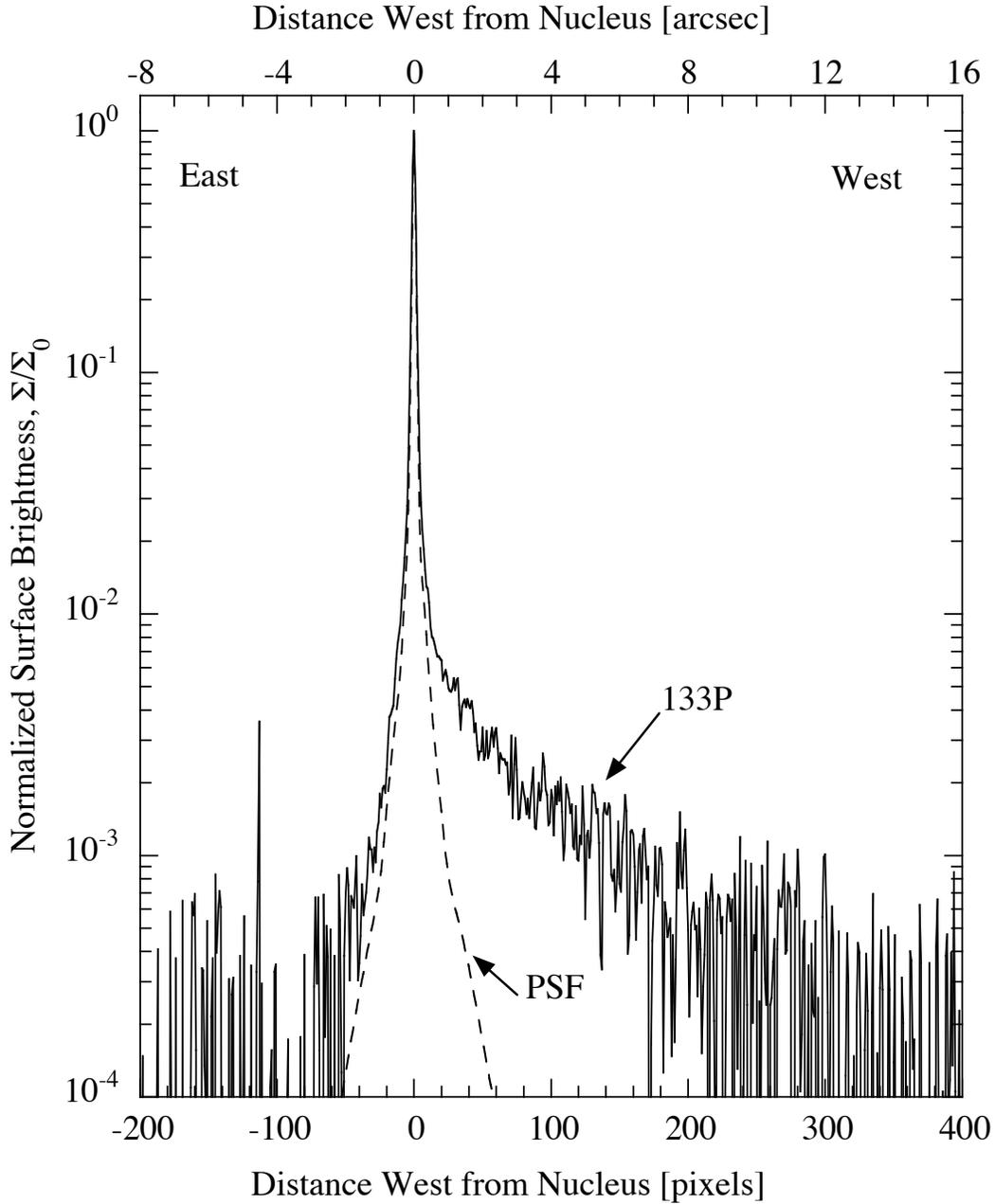}
\caption{Surface brightness of 133P (solid line) and the PSF (dashed line) shown as a function of distance from the nucleus.  The surface brightnesses are the average within a 41 pixel (1.64\arcsec) wide strip centered on the tail.  The peak surface brightness corresponds to $\Sigma_0$  =  15.80 mag.~(arcsec)$^{-2}$.  Negative coordinates are east of the nucleus.   \label{SB_Trail}
} 
\end{center} 
\end{figure}

\begin{figure}
\epsscale{0.8}
\begin{center}
%\plotone{Figure_5.ps}
%\includegraphics[width=0.8\textwidth, angle =270 ]{Figure_5.pdf}
\includegraphics[width=0.8\textwidth]{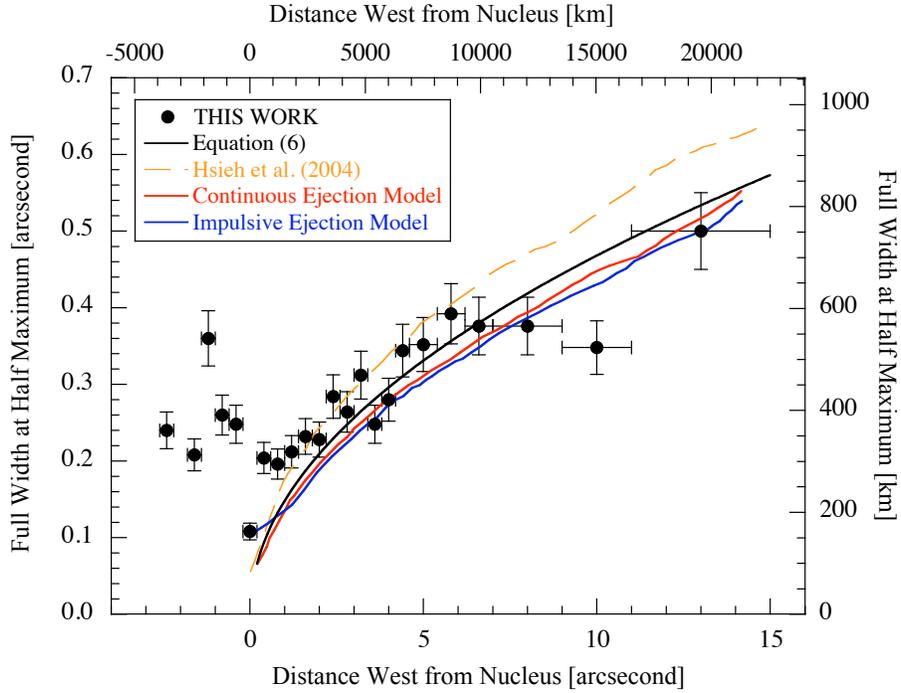}
\caption{Full width at half maximum of the dust tail as a function of distance measured west from the nucleus. Vertical error bars show the estimated $\pm$10\% uncertainties in FWHM.  Horizontal bars mark the range of distances over which each FWHM was measured, increasing as the tail grows fainter. The black circles show measurements from the 2013 July 10 HST data.  The model of Hsieh et al.~(2004) is marked as an orange, dashed line. The red curve shows continuous emission from the date of observation to  $t_0$ = 150 days before, with $q$ = 3.25 and $\beta_{min}$ = 5$\times$10$^{-5}$.  The blue curve shows impulsive emission occurring 70 days before the HST observation, with $q$ = 3.15 and $\beta_{min}$ = 10$^{-4}$.  \label{width}
} 
\end{center} 
\end{figure}

We extracted the surface brightness along the tail  within a 41 pixel (1.64\arcsec) wide rectangular box aligned parallel to the tail axis.  Sky subtraction was obtained from a linear interpolation of the background in equal-sized regions above and below the tail.  The surface brightness profile, shown in Figure (\ref{SB_Trail}), is normalized to a peak value $\Sigma_0$ = 15.80 V mag.~(arcsec)$^{-2}$.  The surface brightness as a function of distance, $d$, along the tail is consistent with $\Sigma \propto d^\gamma$, where $\gamma$=-0.86$\pm$0.06 and 30 $\le d \le$ 200 pixel  (1.2 $\le d \le$ 8.0\arcsec).  The measurements become unreliable beyond about 16\arcsec~west of the nucleus, owing to a combination of extreme faintness and overlap by the images of imperfectly-removed field galaxies (c.f.~Figure \ref{image}). The surface brightness  drops precipitously to the east of the nucleus but, as noted above, dust is evident within a few arcseconds as Feature ``C'' (Figure \ref{image}). 

We measured the ratio of the light scattered from the dust to that scattered from the nucleus.  Given that the nucleus and dust have the same albedo and phase function, this brightness ratio is equivalent to the ratio of the dust cross-section to the nucleus cross-section, $C_d/C_n$.  We find $C_d/C_n$ = 0.36, with an uncertainty that  is dominated by systematics of the data and is difficult to estimate, but is unlikely to be larger than 50\%.   With nucleus cross-section $C_n$ = 15$\pm$2 km$^2$,  we find $C_d$ = 5.4 km$^2$, good to within a factor of two.  We will use this cross-section to estimate the dust mass in Section \ref{discussion}.

\subsection{Dynamical Tail Models}
\label{details}
To advance beyond the order of magnitude considerations in Section \ref{OOM}, we
created model  images of 133P using a Monte Carlo dynamical procedure developed in Ishiguro et al.~(2007) and Ishiguro et al.~(2013). 
It is not possible to obtain unique solutions for the dust properties from the models, given the large number of poorly constrained parameters in the problem.  However, the numerical models are useful because they can help reveal implausible solutions, and identify broad ranges of dust ejection parameters that are compatible with the data.

The dynamics of dust grains are  determined both by their ejection
velocity and by the radiation pressure acceleration and solar gravity, parametrized through their ratio, $\beta$. For spherical
particles, 
$\beta =$  0.57/$\rho a_{\mu m}$, where $\rho$ is the dust mass density in
g cm$^{-3}$ (Burns et al.~1979).  In the rest of the paper, we assume the nominal density $\rho$ = 1 g cm$^{-3}$ and compute particle radii in microns, $a_{\mu m}$, accordingly. We assume that dust particles are ejected symmetrically
with respect to the Sun--comet axis in a cone-shape distribution with half-opening
angle $\omega$ and that the ejection speed is a function of the particle size and, hence, of $\beta$. We adopt the following function for the terminal speed:  

\begin{equation}
V_T = v V_1~\beta^{u_1},
\label{vel}
\end{equation}

\noindent
where $V_1$ is the average ejection velocity of  particles having $\beta$=1.
As considered in Ishiguro et al. (2007), $V_T$ also has a heliocentric distance dependence.   Here, we neglect this dependence because the eccentricity of 133P, $e$=0.16, is small and the $R$ dependence of $V_T$ is weak. The power index, $u_1$, characterizes the
$\beta$ (i.e. size) dependence of the ejection velocity. In Equation (\ref{vel}), $v$ is a random
variable in which the probability of finding $v$ in the range $v$ to $v + dv$ is given by the Gaussian density function,
$P(v)dv$, 

\begin{equation}
P(v) dv = \frac{1}{\sqrt{2\pi} \sigma_v}\exp \left[-
	 \frac{(v-1)^2}{2\sigma_v^2}\right] dv , 
\label{Pv}
\end{equation}

\noindent
where $\sigma_v$ is the standard deviation of $v$. Two thirds of the values fall within $\pm$1$\sigma_v$ of the mean. A power-law size distribution
with an index $q$ was used. The dust production rate at given size 
and time is written:

\begin{equation}
N(a_{\mu m};t)~da = N_0 ~a_{\mu m}^{-q}
 R^{-k} ~da, 
\label{Number}
\end{equation}

\noindent
in the size range of $a_{min} \le a_{\mu m} \le$ $a_{max}$, where
$a_{min}=0.57/ \rho \beta_{max}$ and $a_{max}=0.57/\rho \beta_{min}$, respectively. The power index, $k$, defines the dust production rate as  a function of the heliocentric distance $R$, and $R$ is calculated as a function of time $t$.  In Equation (\ref{Number}), $R$ is expressed in AU.  

Model images were produced by Monte Carlo simulation by solving Kepler's equation including solar gravity and radiation pressure. We
calculated the positions of dust particles on UT 2013 July 10 under conditions given by Equation (\ref{vel})--(\ref{Number}), and derived
the cross-sectional areas of dust particles in the HST/WFC3 CCD coordinate system by integrating with respect to time and particle size, that is,

\begin{equation}
C(x,y) = \int_{t_0}^{t_1} \int_{a_{min}}^{a_{max}} N_{cal}(a_{\mu m},t,x,y)
 ~\pi a_{\mu  m}^2 ~da_{\mu m}~dt
\label{CrossSection}
\end{equation}

\noindent
where $N_{cal}(a_{\mu m},t,x,y)$ is the number of dust particles projected within a
pixel at coordinates (x,y) in a WFC3 CCD image. The model assumes that dust is ejected uniformly over the  interval from $t_0$ to $t_1$, such that $t_0$ is the time elapsed between the start of dust ejection  and our HST observation, and $t_1$ is the time elapsed between the end of dust ejection and our HST observation. 

%%%
The HST data offer four key properties with which to constrain our dust models:

\begin{enumerate}

\item The position angle of the tail is $\theta_{PA}$ =  247.1\degr$\pm$0.4\degr~on UT 2013 July 10.

\item No gap can be discerned between the nucleus and the dust tail.  Any such gap larger than $\sim$0.4\arcsec~would be easily detected.

\item The surface brightness profile of the tail can be approximately represented by $\Sigma \propto d^\gamma$, where 
$\gamma$ = -0.86$\pm$0.06 and 30 $\le d \le$200 pixel  (1.2 $\le d \le$ 8.0\arcsec; c.f.~Figure \ref{SB_Trail}).

\item The full width at half maximum of the antisolar tail is very small, rising from 0.2\arcsec~near the nucleus to 0.5\arcsec~12\arcsec~to the west (c.f.~Figure \ref{width}). 

\end{enumerate} 
%%%%%

\begin{figure}
\epsscale{0.9}
\begin{center}
\plotone{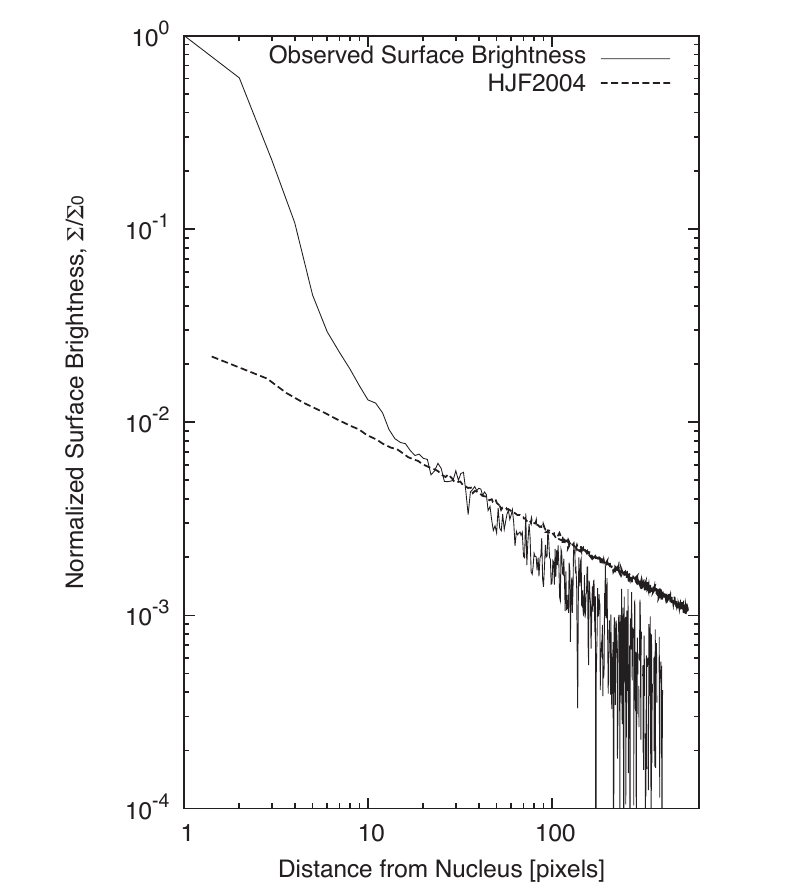}
\caption{Surface brightness of 133P (solid line) shown as a function of distance from the nucleus.  A model computed using dust parameters derived in Hsieh et al.~(2004) is shown as a dashed line.  \label{Surface_HJF2004}
} 
\end{center} 
\end{figure}

As a starting point, we used the model parameters derived by Hsieh et al.~(2004). They assumed $u_1$=1/2,
appropriate for gas-driven dust ejection and a power-law differential size distribution with an index of $q$ =3.5
in the range 0.05 $<$ $\beta$ $<$ 0.5. Hsieh et al.~(2004)
noticed that the tail width was controlled by the particle ejection
velocity perpendicular to the orbital plane, $V_1$, finding $V_1$ = 1.1 m s$^{-1}$. We examined the width and
surface brightness to compare with our observational results assuming
$\omega$ = 90\degr~and $V_1$ = 1.5 m s$^{-1}$. We arbitrarily chose  $t_0$ = 1177 days (time of the last aphelion passage) and $t_1$=0.  Figure (\ref{width}) shows that the Hsieh et al.~(2004) model parameters approximately match the dust trail width in the HST data.  However, Figure \ref{Surface_HJF2004} shows that the model fails to match the new surface brightness profile. 

To try to improve the fit, we examined which parameters in our model most affect the observed quantities. In our model the tail width is largely controlled by  $V_1 \sin w$.   The position angle of the dust tail is largely a measure of the timing and duration of dust ejection ($t_0$ and $t_1$).    The surface brightness distribution depends on the starting time and duration of dust ejection ($t_0$ and $t_1$), the size range of the particles ($\beta_{min}$ and $\beta_{max}$) and the size distribution index  ($q$).  When dust particles are ejected at constant rate over a long interval (specifically, for a time longer than the time needed for the slowest dust particles to travel the length of the tail), a steady-state flow of dust particles results in a surface brightness distribution with $\gamma$ = -0.5 (Jewitt and Meech 1987).  On the other hand, short-term dust supply creates a steeper surface brightness distribution, i.e.~$\gamma < -0.5$, because larger, slower particles are bunched-up near the nucleus increasing the surface brightness there.

A key question is whether dust emission from 133P was impulsive (as would be expected from an impact, for example) or continuous (more consistent with sublimation-driven mass-loss).  We computed two families of models to attempt to address this question.  In the continuous ejection models (CM), the dust ejection is supposed to be steady in the interval from the start time, $t_0$, to the end time, $t_1$, and we assume $t_1$ = 0, corresponding to dust emission continuing up to the epoch of observation.  In the impulsive ejection models (IM), we limit the dust ejection to a single
day (i.e.~dust is released from $t_0$ to $t_1$=$t_0$-1) to approximate an impulse.  We created simulation
images using a wide range of parameters as listed in Table (\ref{parameter}). 

\begin{figure}
\epsscale{0.9}
\begin{center}
%\plotone{Figure_7.ps}
%\includegraphics[width=0.75\textwidth, angle =270 ]{Figure_7.pdf}
\includegraphics[width=0.75\textwidth]{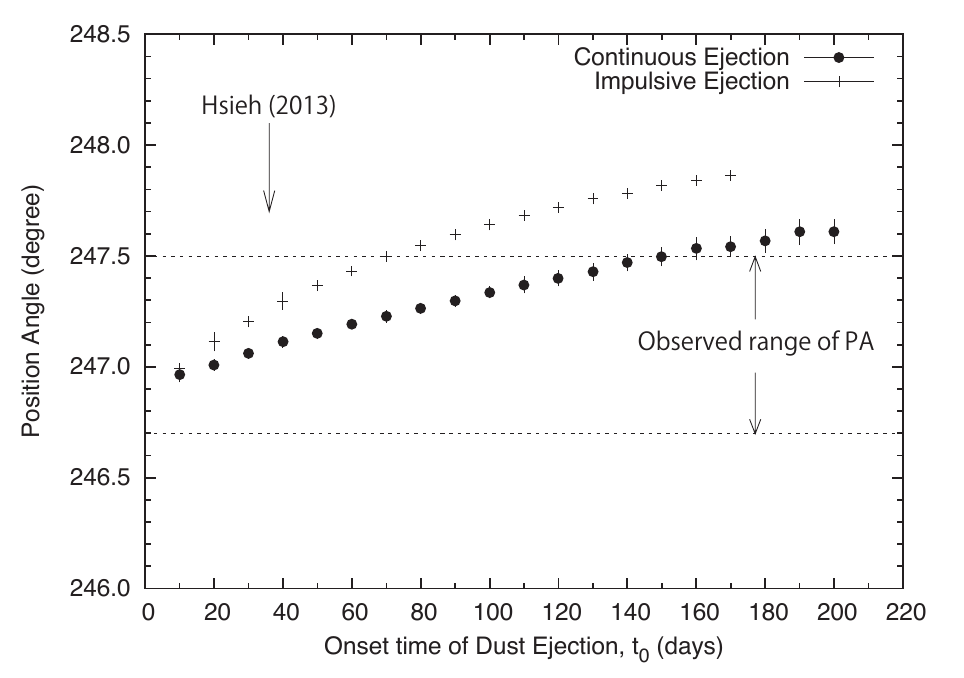}
\caption{Tail position angles as a function of the onset time of dust ejection, $t_0$ (see Equation \ref{CrossSection}). The observed
 range of position angle is shown by dashed lines, and the time of the
 first detection of 133P tail in 2013 appearance (Hsieh et al.~2013) is shown
 by a vertical  arrow. \label{model_pa}
} 
\end{center} 
\end{figure}

As mentioned above, the tail position angle, $\theta_{PA}$, is largely determined by the ejection epoch. Figure (\ref{model_pa}) shows $\theta_{PA}$ as a function of  $t_0$ for the CM and IM models.  Also shown in the figure is the observed range of tail position angles, measured from the HST data.  We find that, to be consistent with the measured $\theta_{PA}$, continuous ejection must have started within 150 days of the HST observation on July 10 (i.e.~$t_0 \leq  150$ days).  In the case of impulsive ejection, dust release must have occurred within 70 days of the HST observation ($t_0 \leq 70$ days) in order to fit the measured tail position angle.  Separately, the detection of dust on June 4 by Hsieh et al.~(2013), 36 days prior to our Hubble observation, indicates $t_0 >$ 36 days (this is probably a very strong limit to $t_0$ since Hsieh already reported a 50\arcsec~tail on June 4).  We therefore conservatively conclude that 36 $\le t_0 \le$ 150 days and 36 $\le t_0 \le$ 70 days, for the continuous and impulsive dust models, respectively.  The age of the dust is measured in months, not years or days.  Unfortunately, the tail position angle alone does not provide convincing discrimination between the continuous and impulsive ejection models.  
% Hereafter, we imposed conditions of  $36 < t_0  \leq 150$ days for CM
% and $36 < t_0 \leq 70$ days for IM.   

More stringent constraints can be set from the dust trail surface brightness profile, from the absence of a near-nucleus coma-gap and from the measured tail width.  In CM, we found  no model parameter sets that satisfy the observational constraints (1) to (4) when $q \geq$ 3.5 while only 2\% satisfy the constraints when $q=3.0$ is assumed. However, 28\% of the tested parameter sets satisfied these conditions when $q=3.25$; we believe that 3.25 $\le q \le$ 3.5 best represents the size distribution index for the continuous emission models.   For continuous ejection models having $t_0$ = 36 days, we find 2$\times$10$^{-3} \le \beta \le$ 7$\times$10$^{-2}$, corresponding to particle radii 10 $\le a \le$ 300 $\mu$m.  For continuous ejection beginning at $t_0$ = 150 days, we find 6$\times$10$^{-5} \le \beta \le$ 3$\times$10$^{-3}$, corresponding to particle radii 0.2 $\le a \le$ 10 mm.

%For initiation time $t_0$ = 36 days the CM models give $\beta_{min} \le$
%2$\times$10$^{-3}$ while for $t_0$ = 150 days $\beta_{min} \le$  6$\times$10$^{-5}$.   The corresponding particle sizes are $a_{max} \geq$ 0.6 mm to 9.5 mm ($\rho$=1 g cm$^{-3}$ was
%assumed). 

\begin{figure}
\epsscale{0.9}
\begin{center}
%\plotone{Figure_8.ps}
%\includegraphics[width=0.75\textwidth, angle =270 ]{Figure_8.pdf}
\includegraphics[width=0.75\textwidth]{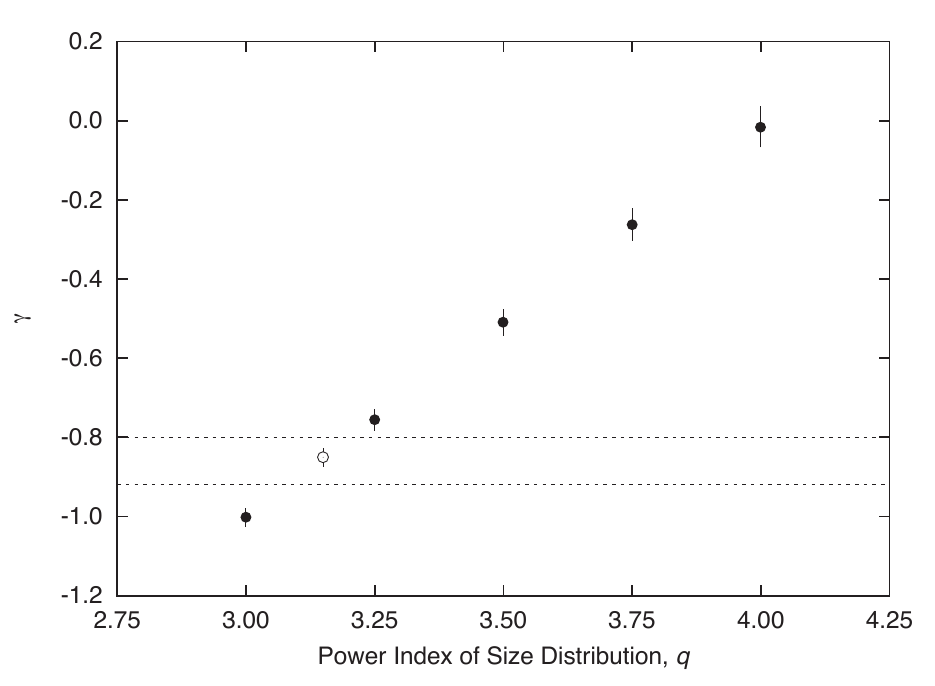}
\caption{Power index of the surface tail brightness distribution
 $\gamma$ vs power 
 index of particle size distribution $q$. Filled circles are derived
 from our model simulations with parameters in Table
 (\ref{parameter}). The dashed lines show the upper and lower limits
 from our observation. Open circle was obtained by the additional
 simulation with $q$=3.15. The errors correspond to 3$\sigma$ of the
 simulation results.
\label{q-gamma}
} 
\end{center} 
\end{figure}

\begin{figure}
\epsscale{0.9}
\begin{center}
\plotone{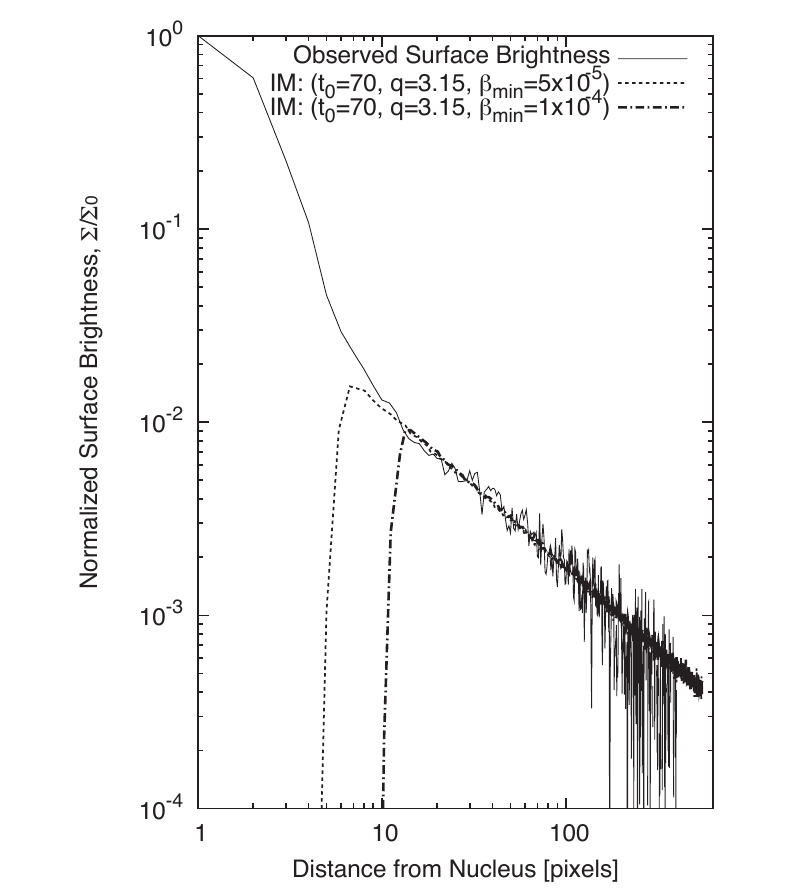}
\caption{Surface brightness of 133P (solid line) and  impulsive models showing the effect of maximum particle size.   
 \label{Surface_IM}
} 
\end{center} 
\end{figure}

The observed tail surface brightness gradient, $\gamma$,  correlates strongly with $q$, as shown in Figure (\ref{q-gamma}). The filled circles denote the average values from our model results with their 3 $\sigma$ uncertainties. The point at $q=3.15$ (open circle in Figure (\ref{q-gamma})) matches the measured surface brightness index, $\gamma$=-0.86$\pm$0.06.   Impulsive models with $q$ = 3.15$\pm$0.05 can match the surface brightness gradient of the tail.  However, large (slow) particles are needed to avoid the growth of a gap between tail and nucleus, as shown in Figure \ref{Surface_IM}.  Assuming that any such gap is smaller than 10 pixel (0.4\arcsec), we estimate $\beta_{min}$ $\leq$ 4$\times$10$^{-4}$ ($t_0$=36) to 1$\times$10$^{-4}$ ($t_0$=70), which correspond to $a_{max} \geq$ 1.4 mm -- 5.7 mm.   Such particles are a factor of ten or more larger than those inferred in earlier work.  These models successfully match the observed brightness
distribution and the measured tail width.  Figure (\ref{width})  compares the observed and model tail widths. 

While the multi-parameter nature of the dust modeling  allows us to find impulsive ejection models to fit the imaging data, the photometry imposes an important, additional constraint. In  the impulsive case, the scattering cross section near the nucleus should decrease with time as a result of radiation pressure sweeping and the absence of a source of dust particle replenishment.   Instead, the coma magnitude in the 0.2\arcsec--1.0\arcsec~annulus shows no evidence for fading  with time 
(Figure \ref{lightcurve}). Furthermore, the HST photometry (Table \ref{photometry}) is consistent with the brightness determined in the June 04 - 14 period by Hsieh et al.~(2013), which itself showed no evidence for fading.  The absence of fading is hard to understand in the context of an impulsive ejection origin, except by highly contrived means (e.g.~ejected particles could brighten by fragmenting in such a way as to offset the fading caused by radiation pressure sweeping).  We conclude that it is very unlikely that the dust ejection was
impulsive.  Protracted emission was independently inferred from observations of 133P in previous orbits (Hsieh et al.~2004, 2010) and is consistent with dust ejection through the production of sublimated ice.

The sunward tail (Feature ``C'' in Figure \ref{difference}) is not present in our basic simulations of 133P because all particles are quickly accelerated to the antisunward side of the nucleus by radiation pressure. We considered two possibilities to explain the existence of the sunward tail. First, high speed dust particles could be launched sunward in a narrow jet (to maintain $V_1 \sin w$, needed to ensure a narrow tail and no resolvable coma).  By experiment we find that a half-opening angle $\omega \ll 5\arcdeg$ would be needed to produce Feature ``C''  in Figure \ref{difference}, but, while this solution is technically possible, it seems contrived.  We favor a second possibility, namely that the sunward tail consists of ultra-large, slow particles ejected during a previous orbit.  Particles released from the nucleus and moving largely under the action of solar gravity will return to the vicinity of the orbit plane every half orbit period, producing a structure known as a ``neck-line'' (Kimura and Liu 1977). Our second possibility is that Feature ``C'' could be 133P's neck-line, produced by the convergence of particles released a full orbit ago.  If so, we estimate that the responsible particles in region C have sizes measured in centimeters or larger.  A simulation of the neck-line structure is shown in Figure (\ref{neckline}).

\section{DISCUSSION}
\label{discussion}
\subsection{Particle Properties: Mass,  Loss Rate and Lifetime}
We estimate the dust mass in 133P as follows.  The mass of an optically thin collection of spheres of individual density $\rho$ and mean radius $\overline{a}$ is related to their total cross-section, $C_d$, by $M_d = 4/3\rho \overline{a} C_d$.  The effective mean radius of the particles contributing to the scattered light depends on the size distribution, $N(a)da$, the scattering area, $\pi a^2$, and the time of residence of particles in the tail. The latter is size-dependent because radiation pressure imparts a velocity to dust particles that scales as $V$ = $V_1/a^{1/2}$, where $V_1$ is a constant. The amount of time that a particle of size $a$ spends in a given pixel of size $\delta x$ is $t$ = $\delta x/V$ = $(K a^{1/2})$, where $K$ is a constant.  Putting these together, we find that the effective mean particle radius weighted by the size distribution, the scattering cross-section and the time of residence is

\begin{equation}
\overline{a} = \frac{\int_{a_{min}}^{a_{max}} a \pi a^2 K a^{1/2} N(a) da}{ \int_{a_{min}}^{a_{max}} \pi a^2 K a^{1/2} N(a) da}
\end{equation}

\noindent with $N(a)$ computed from Equation \ref{Number}.  The continuous ejection models give 3.25 $\le q \le$ 3.5 (Table \ref{parameter}). With these values, and assuming $a_{max} \gg a_{min}$, we obtain  

\begin{equation}
\overline{a} \sim \frac{a_{max}}{5};~ (q = 3.25)
\label{3.25}
\end{equation}

\begin{equation}
\overline{a} \sim \frac{a_{max}}{\ln(a_{max}/a_{min})};~ (q = 3.5)
\label{3.5}
\end{equation}

\noindent For example, in continuous emission starting at $t_0$ = 150 days, with 0.2 $\le a \le$ 10 mm (Table \ref{parameter}), Equations (\ref{3.25}) and (\ref{3.5}) give $\overline{a}$ = 2 to 3 mm.  With $\rho$ = 1000 kg m$^{-3}$, we estimate the mass in the particle tail as $M_d \sim$1.8$\times$10$^7$ kg.   If this mass were released uniformly over  $\sim$150 days, the implied mean mass-loss rate would be $dM_d/dt \sim$ 1.4 kg s$^{-1}$, consistent with previous estimates of the mass-loss rates in 133P.  The corresponding values for the $t_0$ = 36 day solution are $\overline{a}$ = 60 to 90 $\mu$m, $M_d \sim$5$\times$10$^5$ kg and $dM_d/dt \sim$ 0.2 kg s$^{-1}$.  We conclude that the continuous emission models bracket the mass-loss rates in the range 0.2 $\le dM_d/dt \le$ 1.4 kg s$^{-1}$.  These rates are 10 to 100 times larger than found by Hsieh et al.~(2004), largely because they used much smaller mean dust grain sizes than found here.

The mass-loss lifetime of the nucleus 133P can be estimated from

\begin{equation}
\tau = \frac{M_n}{f_o f_a (dM_d/dt)}
\label{tau}
\end{equation}

\noindent in which $M_n$ is the mass of the nucleus, $f_o$ is the fraction of each orbit over which 133P ejects mass and $f_a$ is the duty cycle (ratio of active periods to elapsed time) for mass-loss between orbits.  The mass of a 2 km radius spherical nucleus of density $\rho$ = 1300 kg m$^{-3}$ is $M_n \sim$ 4$\times$10$^{13}$ kg.  Observations show that $f_o \sim$ 0.2 (Toth 2006) while $f_a$ is unknown.    Substituting into Equation (\ref{tau}), and expressing $\tau$ in years, we find 5$\times$10$^6 \le  \tau f_a \le$ 3$\times$10$^7$ yr.  Strong upper limits to the duty cycle are obtained by setting $\tau$ = 4.5$\times$10$^9$ yr, the age of the solar system. Then, we must have $f_a <$ 0.001 for ice in 133P  to  survive for the age of the solar system.  This limit is consistent with most current (albeit widely scattered) observational limits from ensemble asteroid observations, according to which the instantaneous ratio of active to inactive asteroids is  1:6 to 1:60 (Hsieh 2009), $<$1:300 (Hsieh and Jewitt 2006), $<$1:400 (Sonnet et al.~2011), $<$1:25,000 (Gilbert and Wiegert 2010) and $<$1:30,000 (Waszczak et al.~2013).   It has been suggested that 133P might be a product of a recent ($\lesssim$10 Myr) asteroid-asteroid collision (Nesvorny et al.~2009). If so, the preservation of ice would be trivial even for much larger duty cycles.

\subsection{Ejection Mechanism}

The main observation suggesting that sublimation drives mass loss from 133P is the seasonal recurrence of activity.  Episodes of mass loss have repeated in the months near and immediately following perihelion in four consecutive orbits but mass loss was absent in between (Hsieh et al.~2004, Hsieh and Jewitt 2006, Hsieh et al.~2010, Hsieh et al.~2013).  Recurrence at a particular orbital phase strongly suggests a thermal trigger for the activity, and is reminiscent of the orbitally modulated mass loss observed in ``normal'' comets from the Kuiper belt.  None of the other mechanisms considered as drivers of mass loss from the active asteroids (including impact, rotational breakup, electrostatic ejection, thermal fracture and desiccation cracking) can produce seasonal activity on 133P in any but the most contrived way (Jewitt 2012).   Separately, the protracted nature of the emission in 133P is not readily explained by the other mechanisms but it is natural for sublimating ice.  Furthermore, we obtained good model fits to the data when assuming a drag-like size-velocity relation ($V \propto a^{-1/2}$) but not when using a $V$ = constant velocity law.  The multi-parameter nature of the dust modeling problem means that we cannot formally exclude other size-velocity laws, but the success of the drag-like relation is at least consistent with a gas drag origin.  Lastly, Chesley et al.~(2010) reported a 3$\sigma$ confidence detection of non-gravitational acceleration in 133P, most simply interpreted as a rocket effect from asymmetric sublimation.

However, the immediate problem for gas drag is that the solid particles are launched much more slowly than expected.  Micron-sized grains reach only $V_1 \sim$ 1 m s$^{-1}$, as shown both by the order of magnitude approach in Section (\ref{OOM}) and by the detailed models in Section (\ref{details}) (see also Figure \ref{width}).  This is quite different from active comets near the sun, in which micron-sized particles are dynamically well-coupled to the outflowing gas and attain terminal velocities comparable to the local gas velocity while even centimeter-sized grains exceed 1 m s$^{-1}$ (Harmon et al.~2004).    Whipple's (1951) model has become the first stop for estimating comet dust velocities (e.g.~Agarwal et al.~2007).  Applied to 133P, his model predicts the speed of micron-sized particles as $V_1 \gtrsim$ 100 m s$^{-1}$ while millimeter grains should travel at 4.5 m s$^{-1}$, about two orders of magnitude faster than measured.  

To examine the problem of the ejection speed more closely, we consider sublimation from an exposed patch of ice located at the subsolar point on 133P.  We solved the energy balance equation for a perfectly absorbing ice surface exposed at the subsolar point, with 133P located at 2.725 AU.  We included heating and cooling of the surface by radiation and cooling by latent heat taken up as the ice sublimates.  The resulting equilibrium sublimation mass flux is $F_s$ = 4$\times$10$^{-5}$ kg m$^{-2}$ s$^{-1}$, which represents a maximum in the sense that the subsolar temperature is the highest possible temperature on the body.   

The area of sublimating surface needed to supply  mass loss at rate $dM_d/dt$ [kg s$^{-1}$] is given by 

\begin{equation}
\pi r_s^2 = \left(\frac{1}{f_{dg} F_s}\right) \frac{dM_d}{dt}
\label{area}
\end{equation}

\noindent where $r_s$ is the radius of a circle having area equal to the sublimating area and $f_{dg}$ is the ratio of dust to gas production rates.  In earlier work (Hsieh et al. 2004) we assumed $f_{dg}$ =1, as was traditional in cometary studies before long wavelength observations made possible the accurate measurement of dust masses in comets.  However, recent measurements of short-period comets have convincingly shown $f_{dg} >$ 1.  For example, infrared observations of comet 2P/Encke, whose orbit is closest to that of 133P among the short-period comets, give 10 $\le f_{dg} \le$ 30 (Reach et al.~2000).  Values $f_{dg} >1$ are physically possible because the ejected dust, although carrying more mass than the driving gas, travels much more slowly, allowing momentum to be conserved.  We conservatively take $f_{dg}$ = 10 and, with 0.2 $\le dM_d/dt \le$ 1.4 kg s$^{-1}$ use Equation (\ref{area}) to obtain 500 $\le \pi r_s^2 \le$ 3500 m$^2$, corresponding to a circular, sublimating patch on the surface with radius 13 $\le r_s \le$ 33 m.  Evidently, only a very tiny fraction of the nucleus surface  ($r_s^2/r_n^2 \sim$ (1 to 5)$\times$10$^{-5}$) of exposed ice is needed to supply the dust mass loss in 133P.

The tiny size of the active area affects the dust speed, as we now show.  Outflow of the gas into the surrounding half-space vacuum will produce a  wind that exerts a drag force on entrained dust particles.  If outgassing occurred from a point source, the gas density would fall in proportion to $d^{-2}$, where $d$ is the distance from the source.  If outgassing occurred from an infinite plane, the gas density would vary as $d^0$, for the same reason that the surface brightness of an extended light source is independent of the distance from which it is viewed.  The intermediate case, in which the source is neither a point nor an infinite plane, requires a full gas dynamic treatment to determine the iso-density surfaces in the expanding gas, which lies beyond the aims of the present work.  

Instead, we have made a ``small source approximation'' (SSA) model as described in the Appendix.  Its solution is plotted in Figure (\ref{V_vs_a}), where it is compared with the Whipple model.    For both the SSA and the Whipple models, we plot two curves.  The upper curves (solid lines) show the terminal dust speeds in the absence of nucleus gravity while the lower curves (dashed lines) show the full solutions including gravity.   The SSA velocity law has a form similar to Whipple's formula for the speed of an escaping particle, but differs in that the source size, $r_s$, now plays a role in the terminal velocity, with smaller $r_s$ corresponding to lower ejection speeds.  Physically, the difference between our result and Whipple's is the length scale over which gas drag accelerates entrained dust particles.  Whipple assumed that sublimation was uniform from the entire sun-facing hemisphere and so, in his model, the length scale is the radius of the nucleus, $r_n$.  In our model, the length scale is the (much smaller) size of the source, $r_s$.    In simple terms, the difference between the models is the difference between a handgun and a rifle; the latter fires a faster bullet than the former because the longer barrel gives a greater acceleration length over which momentum from the explosive gases is transferred to the projectile.

While the SSA model gives lower speeds than the Whipple model, the predicted grain speeds are still about an order of magnitude larger than those measured in 133P (Figure \ref{V_vs_a}).  We see several possibilities for explaining this difference. One possibility is that $F_s$ is  over-estimated because the ice is not located exactly at the subsolar point.  Ice might also be protected beneath a thin refractory mantle where it would receive only a fraction of the full solar insolation, reducing $F_s$.  Of course, a severe reduction in the sublimation flux would further reduce the critical grain size, $a_c$, magnifying the puzzle of how millimeter and centimeter sized grains can be ejected.

Another possibility is that sublimation proceeds, not from a single patch of  radius $r_s$, but from a large number of  sources  individually small compared to $r_s$, but having a combined area $\pi r_s^2$, such that $V_T$ from Equation (\ref{VT}) is further reduced.   This might be a natural result of progressive mantle growth on an aging ice surface, as larger blocks left behind on the surface restrict sublimation to a network of small, unblocked areas.  By Equation (\ref{VT}), an order of magnitude reduction in dust speed to match the measured values would be produced by sublimation from individual sources of scale $r_s \sim$ 0.2 m.

Yet another possibility is that ejected dust particles leave the surface containing some fraction of attached ice, the anisotropic sublimation of which would exert a reaction (``rocket'') force, propelling the grains laterally out of the gas flow.  Small, spherical dust grains should be essentially isothermal, giving no net force, but an irregular distribution of ice within a porous, aspherical, rotating grain  will in general lead to a net sublimation force even from an isothermal particle.  A very modest ice fraction in each grain would be sufficient to propel the particles laterally over distances $\gtrsim r_s$, especially in 133P where $r_s \sim$ 20 m is so small.  The process would be less important on more active (generally better studied) comets because of their larger source regions.  Unfortunately, the data offer no way to determine whether these effects, working either separately or collectively, occur in 133P.  

It is highly improbable that gas drag acting alone could launch particles from the nucleus into the dust tail with terminal velocities (i.e.~velocities at distances $r \gg r_n$) much smaller than the gravitational escape speed  from the nucleus.  In order to  climb out of the gravitational potential well of the nucleus while retaining a terminal velocity $V_T(a)$, dust grains must be launched from the surface at speed $V$ = $(V_T(a)^2 + V_e^2)^{1/2}$, where $V_e \sim$ 1.8 m s$^{-1}$ is the gravitational escape speed from 133P.  For example, millimeter grains by Equation (\ref{vperp}) have $V_T(a) \sim$ 6 cm s$^{-1}$, from which the above relation gives a launch velocity $V  \sim$ 1.801 m s$^{-1}$ (i.e.~only 0.05\% larger than $V_e$).  Gas drag alone is unable to provide such fine-tuning of the ejection velocity over a wide range of particle sizes.   

The solution to this problem may be that gravity is nearly or completely negated by centripetal acceleration on parts of 133P, as a result of  the elongated shape and 3.5 hr rotation period.   We envision very weak gas flow driven by sublimation near the  tips of the elongated nucleus, perhaps in the form of a fluidized bed on which gravity is reduced nearly to zero.  In this scenario, the particles would accelerate by gas drag according to a $V \propto a^{-1/2}$ velocity-size relation, consistent with the data.  Their escape would be unimpeded by gravity so that even the largest, slowest particles could escape to be swept up by radiation pressure.  Forces other than gas drag might eject particles under these circumstances, but only sublimation can account for the observed seasonal modulation of the activity on 133P, as originally proposed (Hsieh et al.~2004). In this regard, 133P is a C-type object (Hsieh et al.~2004) and, as noted above, must have density $\rho_n \gtrsim$ 1300 kg m$^{-3}$ (Equation \ref{density}) if regolith material is to be gravitationally bound at its rotational extremities.  We note that the average density of C-type asteroids is 1300$\pm$600 kg m$^{-3}$  (Cary 2012), consistent with $\rho_n$.    A hybrid solution in which gas drag provides the force to expel particles with rotation reducing the potential barrier is thus plausible, although not proved.

\clearpage

\section{SUMMARY}

We present new, time-resolved observations of main-belt comet 133P/Elst-Pizarro when in an active state on 2013 July 10.  The new data allow us to examine 133P at unprecedented resolution (Nyquist-sampled resolution is 0.08\arcsec~corresponding to 120 km at 133P).

\begin{enumerate}

\item The principal manifestation of activity in 133P is an anti-sunward tail $>$9$\times$10$^4$ km in length and $<$800 km in width.  The tail consists of dust particles released at low-speeds from the nucleus no earlier than five months prior to the Hubble observation and swept away from it by solar radiation pressure.  A previously unobserved, ultralow surface brightness sunward tail, $\sim$10$^3$ km in linear extent, consists of centimeter-sized and larger particles released in a previous orbit.  There is no resolved coma.

\item The best-fit models involve continuous emission, for periods of months, of dust at rates 0.2 to 1.4 kg s$^{-1}$.  These rates are so small that they could be sustained by the nucleus for billion-year timescales provided the duty-cycle is $\lesssim$ 10$^{-3}$.

\item The main parameters of the nucleus determined from HST data (absolute magnitude $H_V$ = 15.70$\pm$0.10, equivalent circular radius $r_n$ = 2.2$\pm$0.4 km, axis ratio $\sim$1.5:1, semi-axes 2.7 $\times$ 1.8 km and rotation period $\sim$3.5 hr) are compatible with earlier determinations from ground-based data.

\item Characteristic dust speeds are $V \sim$ 1.8 $a_{\mu m}^{-1/2}$ m s$^{-1}$, where $a_{\mu m}$ is the particle radius in microns. Such low speeds are, in part, a consequence of the small source size and resulting limited coupling length on 133P, but additional effects must operate to reduce the dust speeds to the observed level.

%\item The measured properties of 133P (including the seasonal emission of dust on timescales of months, the absence of detected gaseous emission bands,  the low speeds of ejected particles and their conformance to a $V \propto a_{\mu m}^{-1/2}$ velocity law) remain consistent with the hypothesis that mass loss is driven by gas drag from sublimated ice. We cannot prove that this explanation is correct, and no plausible alternative hypothesis has been found.

\item The key properties of 133P are the seasonal emission of dust on timescales of months,  the low speeds of ejected particles, their conformance to a $V \propto a_{\mu m}^{-1/2}$ velocity law and the absence of detected gaseous emission bands.  We suggest a hybrid hypothesis in which dust is accelerated by weak gas flow due to sublimated water ice while dust ejection at sub-escape  speeds is assisted by centripetal acceleration from rapid nucleus rotation.  Acting separately, neither gas drag nor centripetal effects can account for the key properties of 133P.  While the hybrid hypothesis remains conjectural, no plausible alternative has been identified.

\end{enumerate}

\acknowledgments
We thank Henry Hsieh, Pedro Lacerda and the anonymous referee for comments on this manuscript.  We thank Alison Vick, Tamas Dahlem and other 
members of the STScI ground system team for their expert help in planning and scheduling these Target of Opportunity observations.  Based on observations made with the NASA/ESA \emph{Hubble Space Telescope,} with data obtained at the Space Telescope Science Institute (STScI).  Support for program 13005 was provided by NASA through a grant from the Space Telescope Science Institute, which is operated by the Association of Universities for Research in Astronomy, Inc., under contract NAS 5-26555.
Other support was provided by NASA's Planetary Astronomy program.  

\clearpage

\section{Appendix: The Small Source Approximation}

The ``small source approximation'' is a toy model intended to capture the physical essence of a complicated gas dynamics problem. The geometry is shown Figure \ref{diagram}, with a dust particle located at point $P$ along a line connecting the nucleus center with the center of the dust source. Off-axis motions are not considered and we assume that dust speed $V \ll V_g$, the gas speed.   We represent  the source by a surface patch of horizontal radius $r_s$ (c.f.~Equation \ref{area}) and represent the gas density at distance, $d$, above the source by

\begin{equation}
\rho_g(d) = \rho_g(r_n) \left[\frac{r_s}{d + r_s}\right]^2
\end{equation}

\noindent where $\rho_g(r_n)$ is the gas density at the sublimating surface.  This function has the desirable properties that $\rho_g(d) \propto d^{-2}$ for $d \gg r_s$, while, $\rho_g(d) \propto d^{0}$ for  $d \ll r_s$, .

The dust particle is separated from the center of the nucleus by distance $r$.  The nucleus has radius $r_n$ and therefore $d = r - r_n$.    A dust particle of radius $a$ intercepts momentum from gas sweeping past at speed $V_g$ at the rate $C_D \rho_g(d) \pi a^2 V_g^2/2$, where $C_D \sim$ 1 is a dimensionless drag coefficient that depends on the shape and nature (e.g.~compact vs.~fluffy) of the grain.   The mass of the dust grain is just 4/3$\pi \rho a^3$, where $\rho$ is the grain density.  The acceleration of the grain due to gas drag is therefore

\begin{equation}
\alpha = \frac{3 C_D V_g^2 \rho_g(r_n) }{8\rho a}\left[\frac{r_s}{r + r_s - r_n}\right]^2
\end{equation}

In addition, the grain experiences a downward acceleration due to gravity, $-GM_n/r^{-2}$, where $M_n$ is the mass of the nucleus.  The equation of motion can then be written

\begin{equation}
V\frac{dV}{dr} = \frac{3 C_D V_g^2 \rho_g(r_n) }{8\rho a}\left[\frac{r_s}{r + r_s - r_n}\right]^2 - \frac{G M_n}{r^2}
\end{equation}

Integrating from the surface of the nucleus, $r = r_n$ where $V$ = 0, to $r = \infty$, where $V$ = $V_T$, 

\begin{equation}
\int_0^{V_T} VdV = \int_{r_n}^\infty \left(\frac{3 C_D V_g^2 \rho_g(r_n) }{8\rho a}\left[\frac{r_s}{r + r_s - r_n}\right]^2 - \frac{G M_n}{r^2}\right) dr
\end{equation}

Noting that $\rho_g(r_n)$ = $F_s(r_n)/V_g$, and writing $M_n = 4/3\pi \rho_n r_n^3$, where $\rho_n$ is the density of the nucleus, we obtain an expression for the terminal velocity in the SSA:

\begin{equation}
V_T = \left(\frac{3 C_D V_g F_s(r_n) r_s}{4 \rho a} - \frac{8\pi G \rho_n r_n^2}{3}\right)^{1/2}
\label{VT}
\end{equation}

\noindent which is plotted in Figure (\ref{V_vs_a}).  Setting the left hand side of Equation (\ref{VT}) equal to zero gives the critical size

\begin{equation}
a_c = \frac{9 C_D V_g F_s(r_n) r_s }{32 \pi G \rho_n \rho r_n^2 }
\end{equation}

\noindent above which dust particles are too heavy to be ejected by gas drag.  Substituting values appropriate for 133P, we find $a_c \sim$ 0.1 mm.  Again, $C_D$ and the densities $\rho$ and $\rho_n$ are unmeasured and could easily each be wrong by a factor of 2 or more.

\clearpage

\clearpage

\begin{deluxetable}{lccccccc}
%\tabletypesize{\scriptsize}
\tablecaption{Observing Geometry 
\label{geometry}}
\tablewidth{0pt}
\tablehead{
\colhead{UT Date and Time}  & \colhead{$R$\tablenotemark{a}}  & \colhead{$\Delta$\tablenotemark{b}} & \colhead{$\alpha$\tablenotemark{c}}   & \colhead{$\theta_{-v}$\tablenotemark{d}} &   \colhead{$\theta_{\odot}$\tablenotemark{e}}  & \colhead{$\delta_{\oplus}$\tablenotemark{f}}   }
\startdata

2013 July  10d 06h 49m & 2.725 & 2.060  & 18.7 & 248.7 & 246.8  & -0.54\\

\enddata

%% Text for table notes should follow after the \enddata but before
%% the \end{deluxetable}. Make sure there is at least one \tablenotemark
%% in the table for each \tablenotetext.

\tablenotetext{a}{Heliocentric distance, in AU}
\tablenotetext{b}{Geocentric distance, in AU}
\tablenotetext{c}{Phase angle, in degrees}
\tablenotetext{d}{Position angle of the projected negative heliocentric velocity vector, in degrees}
\tablenotetext{e}{Position angle of the projected anti-Solar direction, in degrees}
\tablenotetext{f}{Angle of Earth above the orbital plane, in degrees}

\end{deluxetable}

\clearpage

%%%%%%
\begin{deluxetable}{lccccc}
%\tabletypesize{\scriptsize}
\tablecaption{Photometry 
\label{photometry}}
\tablewidth{0pt}
\tablehead{
\colhead{Image}  & \colhead{UT\tablenotemark{a}}  & \colhead{$t_i$\tablenotemark{b}}  & \colhead{$V_{0.2}$\tablenotemark{c}} & \colhead{$V_{1.0}$\tablenotemark{d}} & \colhead{$\Delta m$\tablenotemark{e}}   }
\startdata
%2009 June 22.222 - 23.YYY & 0.14Y & 1.YY  & 90Y & 120Y$\pm$YY & 120Y  \\
1 & 16h 52m 38s & 348    & 20.362 & 20.319 & 0.043 \\
2 & 17h 00m 34s  & 348   & 20.348 & --           & --         \\
3 & 17h 08m 30s  & 348   &20.372  & 20.271 & 0.101 \\
4 & 17h 17m 02s  & 348   & 20.538 & 20.483 & 0.055 \\
5 & 17h 24m 58s  & 348   & 20.669 & 20.602 & 0.067 \\
6 & 17h 32m 54s  & 252   & 20.766 & 20.701 & 0.065 \\
7 & 18h 28m 18s  & 348   & 20.368 & 20.316 & 0.052 \\
8 & 18h 36m 14s  & 348   & 20.354 & 20.306 & 0.048 \\
9 & 18h 44m 10s  & 348   & 20.387 & 20.337 & 0.050 \\
10 & 18h 52m 42s  & 348 & 20.441 & 20.390 & 0.051\\
11 & 19h 00m 38s  & 348 & 20.521 & 20.465 & 0.056 \\
12 & 19h 08m 34s & 252  & 20.611 & 20.542 & 0.069 \\
\enddata

%% Text for table notes should follow after the \enddata but before
%% the \end{deluxetable}. Make sure there is at least one \tablenotemark
%% in the table for each \tablenotetext.

\tablenotetext{a}{UT start time of the integration on 2013 July 10}
\tablenotetext{b}{Integration time, seconds}
\tablenotetext{c}{Apparent V magnitude within 5 pixel (0.2\arcsec) radius aperture}
\tablenotetext{d}{Apparent V magnitude within 25 pixel (1.0\arcsec) radius aperture}
\tablenotetext{e}{$\Delta m$ = $V_{0.2} - V_{1.0}$}

\end{deluxetable}
%%%%%

\clearpage

\begin{table}
  \caption{Dust Model Parameters\tablenotemark{a}}
  \label{parameter}
  \begin{center}
    \begin{tabular}{llll}
\hline
Parameter   & Parameter Range Explored & Best-fit values & Unit\\
\hline
$q$   & 3 to 4 with 0.25 interval  & CM: 3.25 to 3.50\\
      &                          & IM: 3.15$\pm$0.05\\
$t_0$ & 10 to 200 with 10 interval & CM: 36 to 150 & days  \\
      &                             & IM: 36 to 70 & days \\
$t_1$ & CM: $t_1$=0       & fixed & days  \\
      & IM: $t_1$=$t_0$-1 & fixed & days  \\
$\beta_\mathrm{max}$\tablenotemark{b} & 1$\times$10$^{-3}$ to 1\tablenotemark{d}
         & CM: $\geq$7$\times$10$^{-2}$($t_0$=36) to 3$\times$10$^{-3}$($t_0$=150) &  -- \\
 & &  IM: $\geq$3$\times$10$^{-3}$($t_0$=36) to 1$\times$10$^{-3}$($t_0$=70) &  -- \\
$\beta_\mathrm{min}$\tablenotemark{c} & 5$\times$10$^{-6}$ to 5$\times$10$^{-2}$ \tablenotemark{e}
 & CM: $\leq$2$\times$10$^{-3}$($t_0$=36) to 6$\times$10$^{-5}$($t_0$=150) & --\\
 & &  IM: $\leq$4$\times$10$^{-4}$($t_0$=36) to 1$\times$10$^{-4}$($t_0$=70) & --\\
$V_1$ & 1.5 $ \sin^{-1} w$ & fixed & m s$^{-1}$ \\
$u_1$ & 1/2     & fixed &\\
$k$   & -3 & fixed    & -- \\
$\sigma_v$ & 0.3 & fixed & --\\
$w$ & 90  & fixed & degree\\
\hline
%\end{tabular}
    \tablenotetext{a}{~CM: Continuous ejection model, IM, Impulsive ejection model}
   \tablenotetext{b}{~The smallest value of $\beta_{max}$ needed to fit the data}
   \tablenotetext{c}{~The largest value of $\beta_{min}$ needed to fit the data}    
   \tablenotetext{d}{~1/40 of full range in logarithmic space}
   \tablenotetext{e}{~1/30 of full range in logarithmic space}
       \end{tabular}
  \end{center}
\end{table}
\clearpage

\clearpage

\clearpage

\clearpage

\clearpage

\clearpage

\clearpage

\clearpage

\clearpage

\clearpage

\clearpage

\clearpage

\begin{figure}
\epsscale{0.9}
\begin{center}
\plotone{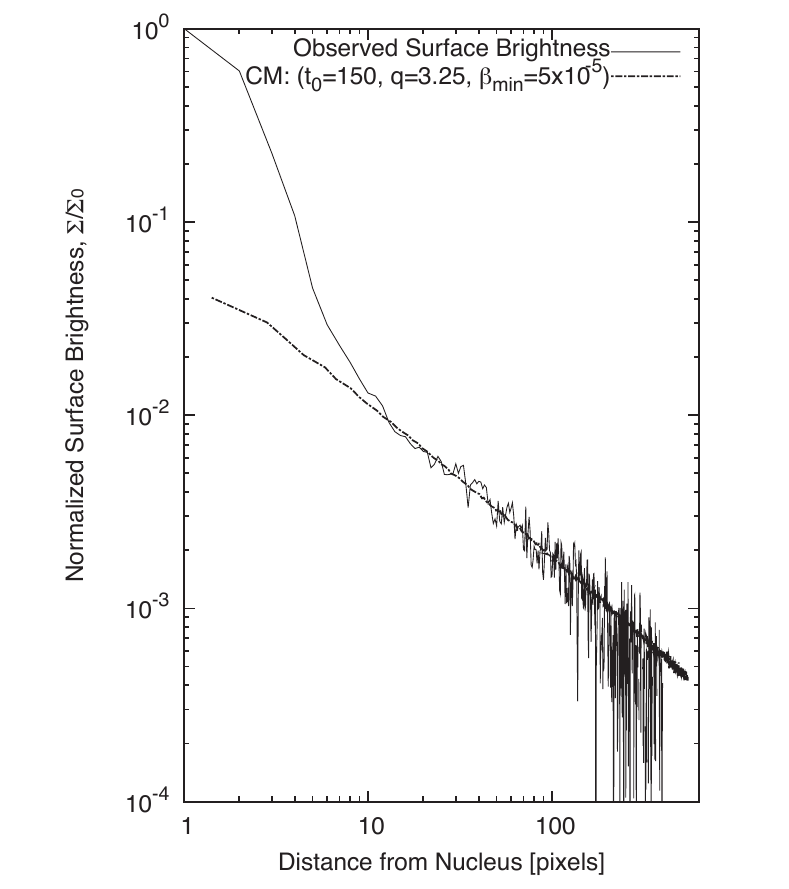}
\caption{Surface brightness of 133P (solid line) and a model in which dust is ejected continuously starting 150 days before the HST observation on July 10.  \label{Surface_CM}
} 
\end{center} 
\end{figure}

\clearpage

\begin{figure}
\epsscale{1.0}
\begin{center}
%\plotone{Figure_11.ps}
%\includegraphics[width=0.75\textwidth, angle =270 ]{Figure_11.pdf}
\includegraphics[width=0.75\textwidth]{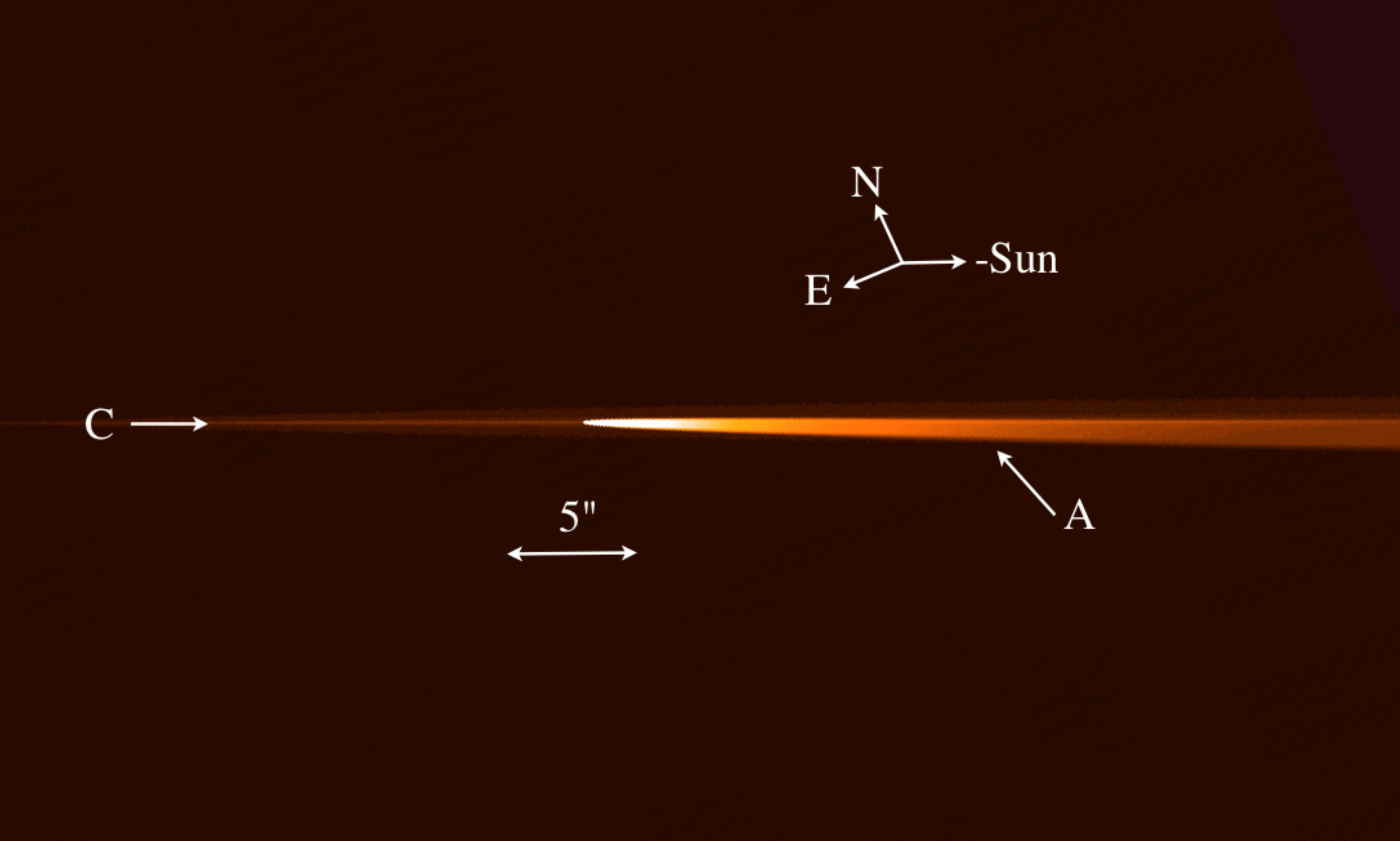}
\caption{Model of neckline structure (C)  (c.f. Figure \ref{difference}).  The neckline structure best seen to the East of the nucleus consists of particles ejected 5.5 years (i.e. one orbit period) before the epoch of the HST observations, with $\beta_{min}$ = 10$^{-5}$.  The main tail ``A'' is also marked.  \label{neckline}
} 
\end{center} 
\end{figure}

\clearpage

\begin{figure}
\epsscale{0.85}
\begin{center}
\plotone{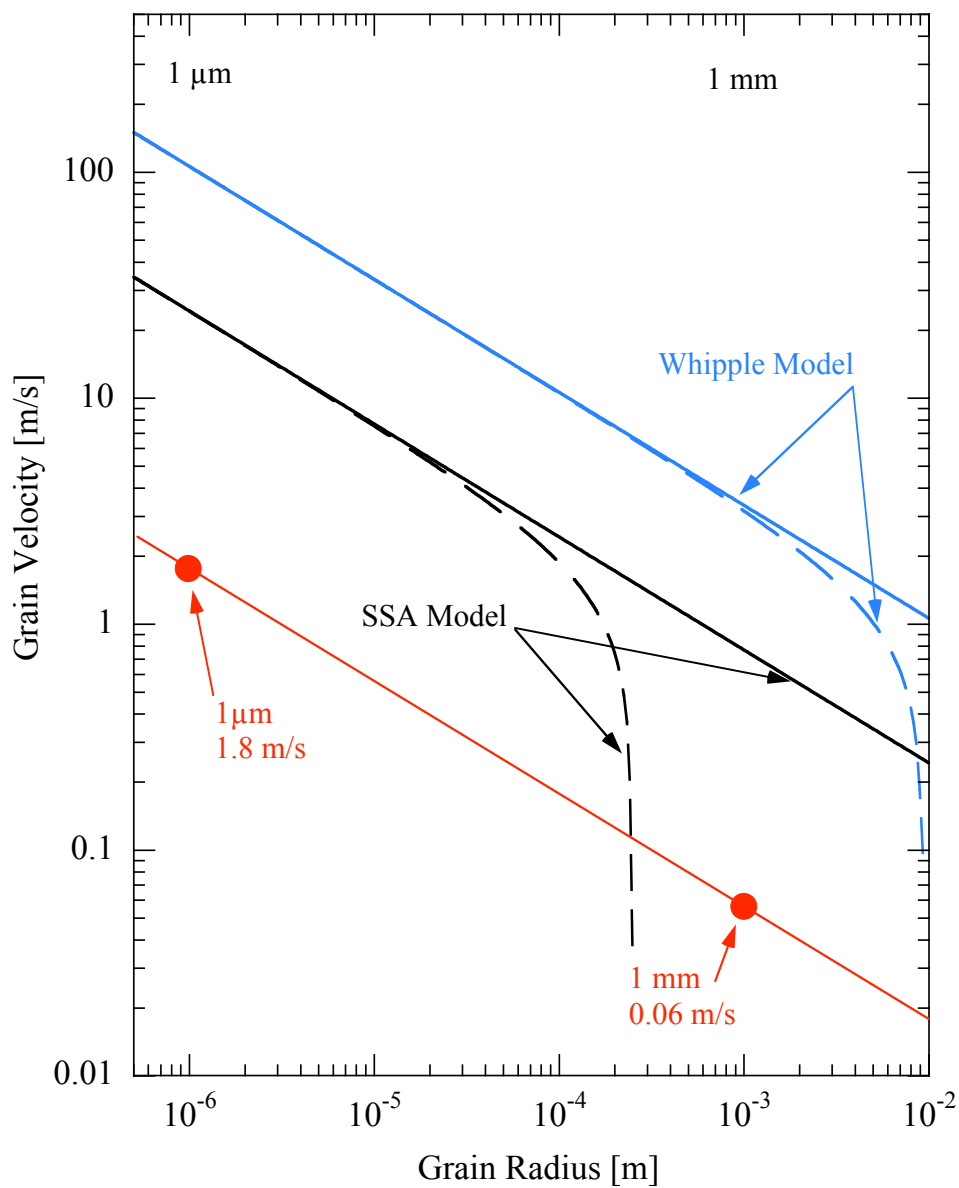}
\caption{Models of the dust grain ejection velocity from 133P as a function of particle radius.  For both SSA (Appendix) and Whipple models, we plot two curves.  The upper (solid) lines show the terminal velocity that would be achieved in the absence of nucleus gravity.  The lower (dashed) lines show the full solutions including gravity.   The red line shows Equation (\ref{vperp}) while the red circles mark speeds of 1 $\mu m$  and 1 mm dust particles measured in 133P.  \label{V_vs_a}
} 
\end{center} 
\end{figure}

\clearpage

\begin{figure}
\epsscale{0.85}
\begin{center}
\plotone{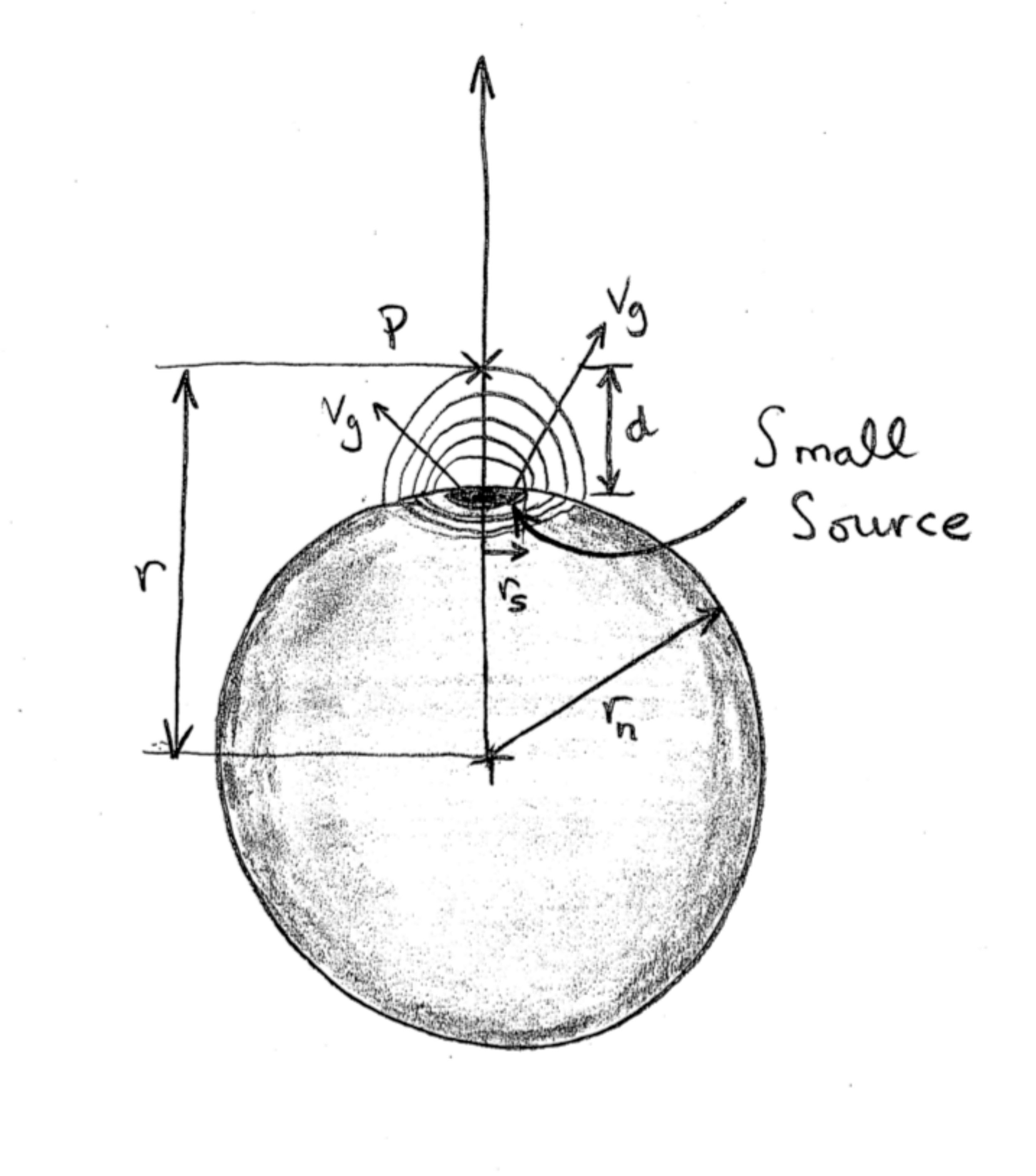}
\caption{Schematic diagram showing the geometry assumed in the Small Source Approximation.  Dust particle $P$ is located at distance $r$ from the center of a spherical nucleus of radius $r_n$, at height $d = r - r_n$ above the surface.  Sublimated gas streams outward at speed $V_g$ from a surface patch of radius $r_s \ll r_n$. See Appendix for a detailed discussion. \label{diagram}
} 
\end{center} 
\end{figure}

\end{document}